\documentclass[lettersize,journal]{IEEEtran}
\usepackage{amsmath,amsfonts}
\usepackage{algorithmic}
\usepackage[ruled]{algorithm2e}
\usepackage{array}
\usepackage[caption=false,font=normalsize,labelfont=sf,textfont=sf]{subfig}
\usepackage{textcomp}
\usepackage{stfloats}
\usepackage{url}
\usepackage{verbatim}
\usepackage{graphicx}
\usepackage{cite}
\usepackage{array}
\usepackage{amssymb}
\usepackage{textcomp}
\usepackage{stfloats}
\usepackage{url}
\usepackage{verbatim}
\usepackage{graphicx}
\usepackage{cite}
\usepackage{cuted}
\usepackage{bm}
\usepackage{calligra}

\usepackage{amsmath} 
\usepackage{booktabs}
\usepackage{epsfig}
\usepackage{epstopdf}
\usepackage{amsmath}
\usepackage{subfig,graphicx}
\usepackage{multirow}
\usepackage{epsfig}
\usepackage{amsmath}
\usepackage{epstopdf}
\usepackage[justification=centering]{caption}
\usepackage{color}
\usepackage{graphicx} 
\usepackage{amssymb}
\usepackage{epsfig}
\usepackage{epstopdf} 
\usepackage{bbm}

\begin{document}

\title{Online Adaptive Real-Time Beamforming Design for Dynamic Environments in Cell-Free Systems}

\author{Guanghui~Chen,~\IEEEmembership{Graduate~Student~Member,~IEEE}, Zheng~Wang,~\IEEEmembership{Senior~Member,~IEEE}, Hongxin~Lin, Pengguang~Du,~\IEEEmembership{Graduate~Student~Member,~IEEE}, Yongming~Huang,~\IEEEmembership{Senior~Member,~IEEE}
	\thanks{This work was supported by the National Natural Science Foundation of China under Grant 62225107, the Natural Science Foundation on Frontier Leading Technology Basic Research Project of Jiangsu under Grant BK20222001, the Fundamental Research Funds for the Central Universities under Grant 2242022k60002. }
	
	\thanks{G. Chen, Z. Wang, P. Du and Y. Huang are with the School of Information Science and Engineering, and the National Mobile Communications Research Laboratory, Southeast University, Nanjing 210096, China; G. Chen and Y. Huang are also with the Pervasive Communications Center, Purple Mountain Laboratories, Nanjing 211111, China. (e-mail:cgh@seu.edu.cn, wznuaa@gmail.com, pgdu@seu.edu.cn, huangym@seu.edu.cn).}
	\thanks{H. Lin is with the Purple Mountain Laboratories, Nanjing 211111, China. (email: linhongxin@pmlabs.com.cn).}
	
}



\maketitle

\begin{abstract}
In this paper, we consider real-time beamforming design for dynamic wireless environments with varying channels and different numbers of access points (APs) and users in cell-free systems. Specifically, a sum-rate maximization optimization problem is formulated for the beamforming design in dynamic wireless environments of cell-free systems. To efficiently solve it, a high-generalization network (HGNet)  is proposed to adapt to the changing numbers of APs and users. Then, a high-generalization beamforming module is also designed in HGNet to extract the valuable features for the varying channels, and we theoretically prove that such a high-generalization beamforming module is able to reduce the upper bound of the generalization error. Subsequently, by online adaptively updating about 3\% of the parameters of HGNet, an online adaptive updating (OAU) algorithm is proposed to enable the online adaptive real-time beamforming design for improving the sum rate. Numerical results  demonstrate that the proposed HGNet with OAU algorithm achieves a higher sum rate with a lower computational cost on the order of milliseconds, thus realizing the real-time beamforming design for dynamic wireless environments in cell-free systems.
\end{abstract}


\section{Introduction}
\IEEEPARstart{R}{ecently}, cell-free systems have received considerable attentions \cite{ref18}, \cite{bib30}.  By connecting all access points (APs) to a central processing unit (CPU) via backhaul links, cell-free systems allow multiple APs to collaboratively design beamforming to serve users within the network coverage area, thus eliminating many interference issues present in cellular systems \cite{ref19}, \cite{bib29}. Nevertheless, beamforming design is a nonconvex optimization problem that is difficult to solve efficiently \cite{ref20},\cite{bib31}. Conventional optimization algorithms like the weighted minimum mean square error (WMMSE) algorithm \cite{ref9} usually use the convex approximation to obtain a locally optimal solution of the beamforming design. Unfortunately, most of them require multiple iterations and matrix inversions, which are difficult to meet the demands of the real-time beamforming design.

To facilitate the real-time implementation, deep learning has been extensively applied to the beamforming design of cell-free systems\cite{bib32},\cite{bib33}. Notably, by applying graph neural networks (GNNs), the work in \cite{ref17} proposed an Edge-GNN to solve the sum rate maximization problem of cooperative beamforming design. Based on the weight sharing mechanism of convolutional neural networks (CNNs), the paper in \cite{ref12} designed a SUNet with high computational efficiency for achieving the downlink beamforming design in cell-free systems. Once trained, these deep learning algorithms require only
simple feed-forward computations to infer beamforming, enhancing the  real-time reflection speed compared to WMMSE algorithm. However, these methods generally assume to work in a fixed configuration, i.e., the channel conditions are the same for the training and the inference phases.

In practice, communication systems normally operate in dynamic wireless environments due to user mobility  and random distributions for various transmission media \cite{ref21}. For example, if a user moves from indoors to  outdoors, or from highly dense places to open places within a period of time, the channel changes accordingly  from Rayleigh fading with non-line-of-sight (NLoS) to Rician fading with LoS \cite{bib38}. Given these  realistically dynamic wireless environments, most deep learning-based beamforming design algorithms struggle to achieve good performance
under real-time requirements. This is since the varying channels cause the data distribution fed into deep learning to be different in the training and the inference phases. Such shift violates the basic assumptions of deep learning, i.e., the same data distribution in the training and the inference stages can yield better generalization performance \cite{bib39}.  Although retraining the model with current channel data can improve the generalization of varying channels, it is time-consuming and insufficient for real-time applications.

There has been some pioneering works  utilizing deep learning to consider dynamic wireless environments, where the channels vary over periods while remaining constant within each period \cite{bib26},\cite{bib27},\cite{bib28}. Specifically, by using continuous learning to adapt to a new period without forgetting the knowledge learned from previous ones, 
\cite{bib26} optimized power allocation for dynamic wireless environments in  single-input
single-output (SISO) cellular systems.  The work in \cite{bib27}  proposed a meta-gating framework including outer and inner networks to realize the beamforming design for dynamic wireless environments in multiple-input single-output (MISO) cellular systems. The outer network evaluated the importance of the inner network’s parameters under varying channels, and then decided which subset of the inner network should be activated through a gating operation. For intelligent omnidirectional surface assisted MISO cellular systems,  \cite{bib28}  proposed a meta-critic reinforcement learning capable of recognizing changes in dynamic wireless environments and automatically performing the self-renewal of the learning mode, while the beamforming design was performed by the low-performance zero-forcing.

Although the above methods have been applied to dynamic wireless environments, it is still a challenge  to apply them to  the real-time beamforming design for dynamic wireless environments in cell-free systems. The reasons are as follows. Firstly, compared to SISO and MISO cellular systems, the multiple-input multiple-output (MIMO) cell-free systems considered in this paper contain numerous  APs and users with multiple antennas, which greatly hinders the application of these methods for the real-time beamforming design. Secondly, compared to the power allocation in \cite{bib26} and the zero-forcing in \cite{bib28}, the number of the optimized variables for the beamforming design is considerably higher, further increasing the challenge of the real-time beamforming design. Thirdly, dynamic wireless environments potentially lead to  a variation in the association relationship between APs and users for  cell-free systems, i.e., the number of  APs and users changes over periods while remaining constant within each period. The algorithms in  \cite{bib26},\cite{bib27},\cite{bib28} probably need to retrain
the network architectures and parameters as the number of users and APs   varies in cell-free systems, which is also difficult to meet real-time requirements.

Motivated by the above challenges, this paper investigates the real-time beamforming design for dynamic wireless environments in cell-free systems.  The objective is to maximize the sum rate while maintaining computational efficiency for the varying channels with the different numbers of APs and users. To achieve the objective, this paper proposes a high-generalization network (HGNet)  with an online adaptive updating (OAU) algorithm. More specifically, the major contributions are summarized as follows: 

\begin{itemize}
	\item[1)] 
	To effectively characterize the dynamic wireless environments in cell-free systems, the channels as well as the number of APs and users are modeled as varying over periods and remaining constant within each period. Meanwhile, a sum-rate maximization optimization problem for the varying channels and the different numbers of APs and users  is built for the beamforming design in dynamic wireless environments of  cell-free systems.
\end{itemize}

\begin{itemize}
	\item[2)] 
	To solve the non-convex optimization problem, we propose  HGNet. Particularly, HGNet incorporates the residual structure of CNNs to map channel state information (CSI) to beamforming with high computational efficiency, which also adapts to the varying numbers of APs and users. HGNet also designs a high-generalization beamforming module to extract the valuable features for the varying channels, and we theoretically prove that the high-generalization beamforming module decreases the upper bound of the generalization error. This helps HGNet to yield a better generalization sum rate  for dynamic wireless environments.
\end{itemize}

\begin{itemize}
	\item[3)] To enable the  online adaptive real-time beamforming design, an OAU algorithm is proposed to  adaptively update about 3\% of the parameters of HGNet online, taking a computationally efficient information entropy as the loss function. This further
	enhances the sum rate performance of dynamic wireless environments with the varying channels and the different numbers of APs and users in cell-free systems.
	
\end{itemize}

\begin{itemize}
	\item[4)]Numerical results are conducted to validate the effectiveness of the proposed HGNet with OAU algorithm.  For  the varying channels and the different numbers of APs and users, the average generalization sum rate performance of HGNet outperforms those of the traditional optimization algorithm WMMSE \cite{ref9}, and the recent deep learning algorithms Edge-GNN \cite{ref17} and SUNet \cite{ref12}, where the average computational cost of HGNet is also the lowest in the order of $10^{-3}$ seconds. Meanwhile, the proposed OAU algorithm further improves the sum rate performance of HGNet with an average computational cost of less than $10^{-2}$ seconds. 	
\end{itemize}

The rest of this paper is organized as follows. In Section \uppercase\expandafter{\romannumeral2}, the system model is introduced, followed by formulating a sum-rate maximization optimization problem with the varying channels and the different numbers of APs and users.  In Section \uppercase\expandafter{\romannumeral3}, the HGNet containing the high-generalization beamforming module is proposed, and a theoretical proof that the high-generalization beamforming module  can reduce the upper bound of the generalization error is given. In Section \uppercase\expandafter{\romannumeral4}, the OAU algorithm is proposed to online adaptively update the parameters of HGNet to realize  the online adaptive real-time beamforming design with increasing
sum rate performance. In Section \uppercase\expandafter{\romannumeral5}, some experimental results for HGNet and OAU algorithm are showed and analyzed. Finally, some
conclusions are provided in Section \uppercase\expandafter{\romannumeral6}.

\textit{Notations:} The scalar, vector, and matrix are denoted by lowercase letter $x$, boldface lowercase letter $\mathbf{x}$, and boldface uppercase letter $\mathbf{X}$, respectively. $\mathbb{C}$ and $\mathbb{R}$ denote the sets of complex and real numbers, respectively. $(\cdot)^H$ denotes the conjugate transpose. $(\cdot)^{-1} $ denotes the matrix inversion. $sup$ denotes the minimum upper bound. 

\section{System Model}

\begin{figure}[t]
	\centering
	\includegraphics[scale=1]{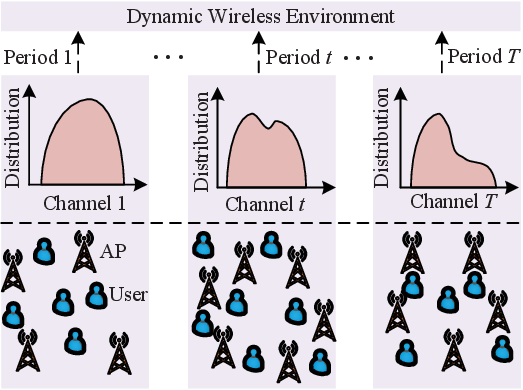}
	\caption{An illustration for dynamic wireless environments with the varying channels and the different numbers of APs and users.}
	\label{fig0}
\end{figure} 

As illustrated in Fig.\ref{fig0}, a dynamic wireless environment for cell-free systems is considered, where the channels as well as the number of APs and users vary between periods and remain constant within each period.  Further, let $\mathcal{T}=\left\{1, \cdots, T\right\}$ denote the set of periods, where $I_t$ users are assumed to access $ Q_t $ APs  in the  $ t^{th} $  period.  $\mathcal{Q}_t=\left\{1_t, \cdots, Q_t\right\}$ and  $\mathcal{I}_t =\left\{1_t, \cdots, I_t\right\}$ denote the sets of APs and users at the $ t^{th} $ period. Each AP and user is configured with $M$ and $N$ antennas, respectively. All APs  are connected to a CPU via  backhaul links for exchanging information, where the CPU can access global CSI to collaboratively design beamforming for improving the system sum rate \cite{ref12}, \cite{ref1}. To simplify the notation,  $ i $ and $ j $ denote the indexes of users, and $ q $ denotes the index of AP. The received signal of the $i^{th}$ user at the $t^{th} $ period is represented as   
\begin{equation}
	\mathbf{y}_{i,t}=\mathbf{H}_{i,t} \mathbf{v}_{i,t} s_{i,t}+\sum_{j \neq i} \mathbf{H}_{i,t} \mathbf{{v}}_{j,t} s_{j,t}+\boldsymbol{z}_{i,t} \in \mathbb{C}^{N\times 1},
	\label{eq1}
\end{equation}
where $ \mathbf{H}_{i,t} \in \mathbb{C}^{N\times Q_tM}$  and  $\mathbf{v}_{i,t}\in \mathbb{C}^{Q_tM\times 1}$
denote the CSI matrix and beamforming vector of the AP set $ \mathcal{Q}_t $ to the $i^{th}$ user at the $t^{th} $ period.  $s_{i,t}$ denotes the data sent to the $i^{th}$ user at the $t^{th} $ period. $\boldsymbol{z}_{i,t} \sim \mathcal{C} \mathcal{N} \left(\mathbf{0}, \sigma_{i,t}^{2} \boldsymbol{\textbf{I}}\right)$ denotes the additive noise. From Eq.(\ref{eq1}), the  achievable rate of the $i^{th}$ user at the $t^{th} $ period  is denoted as
\begin{equation}
	\begin{aligned}
	&{R}_{i,t}=\\&\log \left|\mathbf{I}+\mathbf{H}_{i,t} \mathbf{v}_{i,t}\mathbf{v}_{i,t}^{{H}}\mathbf{H}_{i,t}^{{H}}\left( \sum\limits_{j \neq i} \mathbf{H}_{i,t} \mathbf{v}_{j,t}\mathbf{v}_{j,t}^{{H}}\mathbf{H}_{i,t}^{{H}}+\sigma_{i,t}^{2} \mathbf{I}\right) ^{-1}\right|.
	\label{eq2}
	\end{aligned}
\end{equation}

To better depict the influence brought by dynamic wireless environments, we consider the sum rate maximization problem for the varying channels and the different numbers of APs and users in cell-free systems, i.e.,
\begin{equation}
	\begin{split}
		\underset{\mathbf{v}_{i,t} }{\textrm{max}}&\ \sum_{i=1}^{I_t}{R}_{i,t}
		\\\textrm{s.t.}&\quad \sum_{i=1}^{I_t} \mathbf{v}_{i,t}^{q,{H}}\mathbf{v}_{i,t}^{q}\leq \textrm{P}_{\textrm{max}},\forall q\in\mathcal{Q}_t,\forall t\in\mathcal{T},\\&	D\left(\mathbf{H}_t \right) \neq D\left(\mathbf{H}_{t^\prime} \right), \forall (t\neq t^\prime)\in\mathcal{T},\\&	\left(Q_t,I_t\right)\neq \left(Q_{t^\prime}, I_{t^\prime} \right), \forall (t\neq t^\prime)\in\mathcal{T},
	\end{split}
	\label{eq3}
\end{equation}
where $\textrm{P}_{\textrm{max}}$  is the maximum  transmit power of the AP. $ \mathbf{v}_{i,t}^{q}\in \mathbb{C}^{M\times 1} $ is the beamforming vector of the $q^{th}$ AP to the $i^{th}$ user at the $t^{th} $ period, and  $\mathbf{v}_{i,t}=\left[\mathbf{v}_{i,t}^{1,{H}}, \cdots, \mathbf{v}_{i,t}^{q,{H}} \cdots,\mathbf{v}_{i,t}^{Q_t,{H}} \right]\in \mathbb{C}^{Q_tM\times 1}$.  $\mathbf{H}_t=\left[\mathbf{H}_{1,t}^{{H}}, \cdots, \mathbf{H}_{i,t}^{{H}},  \cdots, \mathbf{H}_{I_t,t}^{{H}}\right]^{{H}} \in \mathbb{C}^{I_tN \times Q_tM}$ denotes the CSI matrix of the AP set $ \mathcal{Q}_t $ to the user set $ \mathcal{I}_t $ at the $t^{th} $ period. $ D\left(\mathbf{H}_t \right)  $ and $ D\left(\mathbf{H}_{t^\prime} \right) $ denote the distribution of $ \mathbf{H}_t $ and $ \mathbf{H}_{t^\prime} $, respectively. 

\emph{Remark 1: The constraint $ D\left(\mathbf{H}_t \right) \neq D\left(\mathbf{H}_{t^\prime} \right), \forall (t\neq t^\prime)\in\mathcal{T} $ in the optimization problem (\ref{eq3}) describes that the channels vary between periods and remain constant within each period for dynamic wireless environments in cell-free systems. Due to the variations in the channels, the input CSI of deep learning has different distributions during  the training and the  inference phases. This violates the underlying assumption of deep learning that better generalization performance is only produced by using the same data distribution in the training and the inference stages. As a result, the constraint $ D\left(\mathbf{H}_t \right) \neq D\left(\mathbf{H}_{t^\prime} \right), \forall (t\neq t^\prime)\in\mathcal{T} $ severely increases the difficulty of solving the optimization problem (\ref{eq3}) leveraging deep learning.}

\emph{Remark 2: The constraint  $ \left(Q_t,I_t\right)\neq \left(Q_{t^\prime}, I_{t^\prime} \right), \forall (t\neq t^\prime)\in\mathcal{T} $ in the optimization problem (\ref{eq3}) describes that   the number of APs and users varies between periods and remains constant within each period for dynamic wireless environments in cell-free systems. The changing numbers of APs and users result in a variation in the dimension of the input CSI for deep learning, requiring the corresponding output dimension of the beamforming to adjust accordingly. In other words, $ \left(Q_t,I_t\right)\neq \left(Q_{t^\prime}, I_{t^\prime} \right), \forall (t\neq t^\prime)\in\mathcal{T} $ requires that the dimension of the output beamforming of deep learning flexibly varies with that of the input CSI. This may require deep learning method to retrain the network architectures and parameters  in a remarkably time-consuming manner, which fails to meet the demands of the real-time beamforming design.}

Note that the optimization problem (\ref{eq3}) is non-convex, which can be solved approximately by traditional optimization algorithms with high computational complexity. However, it is difficult to realize the real-time beamforming design. Deep learning is a good alternative to improve the computational efficiency. Unfortunately, according to \emph{Remark 1} and \emph{Remark 2}, applying deep learning to solve  the optimization problem (\ref{eq3}) is also a challenge in terms of generalization and real-time performance due to the constraints $ D\left(\mathbf{H}_t \right) \neq D\left(\mathbf{H}_{t^\prime} \right), \forall (t\neq t^\prime)\in\mathcal{T} $ and  $ \left(Q_t,I_t\right)\neq \left(Q_{t^\prime}, I_{t^\prime} \right), \forall (t\neq t^\prime)\in\mathcal{T} $. 
As a result, it is necessary to carry out a high-generalization real-time beamforming design for dynamic wireless environments with the varying channels and the different numbers of APs and users in cell-free systems.

\section{Proposed HGNet}
In this section, we pay our attention on designing a HGNet to implement the high-generalization real-time beamforming design for dynamic wireless environments with the varying channels and the different numbers of APs and users in cell-free systems by solving the optimization problem (\ref{eq3}). As illustrated in Fig.2, the proposed HGNet mainly includes an input module, a convolution unit $\mathcal{C}\left(\cdot, \theta_l\right)$, a high-generalization beamforming module $\mathcal{G}\left(\cdot, \theta_l\right)$ and an output module. Specifically, the input module transforms a complex-valued CSI into a real-valued CSI. $\mathcal{C}\left(\cdot, \theta_l\right)$ maps the real-valued CSI  to beamforming with low computational complexity and satisfying $ \left(Q_t,I_t\right)\neq \left(Q_{t^\prime}, I_{t^\prime} \right), \forall (t\neq t^\prime)\in\mathcal{T} $ in the optimization problem (\ref{eq3}). Subsequently,  $\mathcal{H}\left(\cdot, \theta_l\right)$ is specially designed to improve the generalization sum rate performance for $ D\left(\mathbf{H}_t \right) \neq D\left(\mathbf{H}_{t^\prime} \right), \forall (t\neq t^\prime)\in\mathcal{T} $ in the optimization problem (\ref{eq3}). Finally, the output module yields the complex-valued beamforming that satisfies the power constraint.

\begin{figure*}[t]
	\centering
	\includegraphics[width=18.3cm,height=10cm]{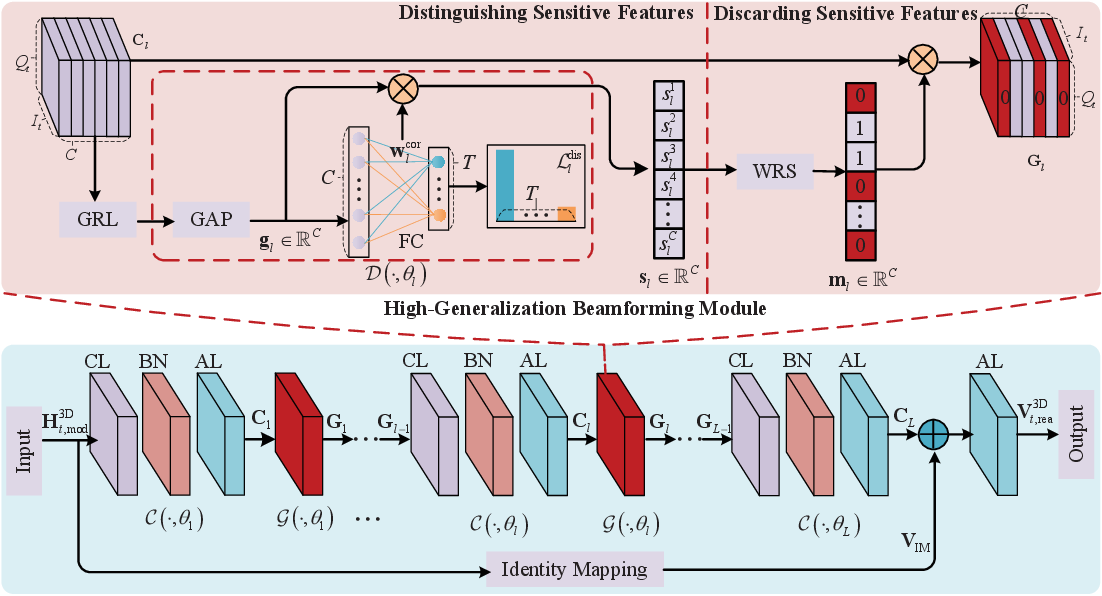}
	\caption{Proposed HGNet.}
\end{figure*}

\subsection{Input}
Since deep learning methods such as CNNs normally deal with three-dimensional (3D) real numbers, the input module converts  $\mathbf{H}_t\in \mathbb{C}^{I_tN \times Q_tM}$ into a 3D real-valued CSI tensor $\mathbf{H}_{t,\text{mod}}^{\text{3D}}\in \mathbb{R}^{Q_t \times I_t \times MN}$ by computing modulus values and dimension transformations.  \footnote{In this paper, the first, second and third dimensions of a 3D tensor are denoted as width, height and third dimension, respectively.}

\subsection{Convolution Unit $\mathcal{C}\left(\cdot, \theta_l\right)$}
$\mathcal{C}\left(\cdot, \theta_l\right)$  aims to achieve the mapping from  $\mathbf{H}_{t,\text{mod}}^{\text{3D}}\in \mathbb{R}^{Q_t \times I_t \times MN}$ to beamforming with fulfilling $ \left(Q_t,I_t\right)\neq \left(Q_{t^\prime}, I_{t^\prime} \right), \forall (t\neq t^\prime)\in\mathcal{T} $ in the optimization problem (\ref{eq3}). Particularly,  $ \left(Q_t,I_t\right)\neq \left(Q_{t^\prime}, I_{t^\prime} \right), \forall (t\neq t^\prime)\in\mathcal{T} $ requires that the dimension of the output beamforming of deep learning should flexibly change with the dimension of the input $\mathbf{H}_{t,\text{mod}}^{\text{3D}}\in \mathbb{R}^{Q_t \times I_t \times MN}$. Fortunately, the
output dimension of the convolution unit is determined
by the input data dimension and the convolutional architectural parameters. Accordingly, it is encouraged to apply the convolution unit to derive some architectural
conditions for satisfying $ \left(Q_t,I_t\right)\neq \left(Q_{t^\prime}, I_{t^\prime} \right), \forall (t\neq t^\prime)\in\mathcal{T} $. On the other hand, the unique weight sharing mechanism of the convolution unit significantly reduces the computational complexity of neural networks, which is also in line with the goal of the real-time beamforming design. Consequently, $\mathcal{C}\left(\cdot, \theta_l\right)$ selects the convolution unit to cascade into a deep structure with $L$ layers, where each layer contains a convolution layer (CL), a batch normalization (BN) layer, and an activation layer (AL). Formally,  the formula of the  $l^{th}$ layer $\mathcal{C}\left(\cdot, \theta_l\right)$ is defined as
\begin{equation}
	\mathbf{C}_{l}=\text{AL}\left(\text{BN}\left(\text{CL}\left(\mathbf{G}_{l-1}, \theta_l\right)\right)\right),
	\label{eq4}
\end{equation}
where $\mathbf{C}_{l}$ denotes the output of $\mathcal{C}\left(\cdot,\theta_l \right) $, and $\theta_l$ is the parameters of HGNet.  $\mathbf{G}_{l-1}$ denotes the input of $\mathcal{C}\left(\cdot,\theta_l \right) $, and $\mathbf{G}_{0}=\mathbf{H}_{t,\text{mod}}^{\text{3D}}\in \mathbb{R}^{Q_t \times I_t \times MN}$. $\text{CL}\left(\cdot,\cdot \right) $ denotes the convolution operation. $\text{BN}\left(\cdot \right) $ denotes the BN operation, which is added after the CL  for reducing the overfitting probability\cite{ref4}.  $\text{AL}\left(\cdot \right) $ denotes the AL operation \cite{ref5}, which selects the commonly used  $ \text{ReLU}(x)=\max (0, x) $. Note that the last layer $\mathcal{C}\left(\cdot,\theta_L \right) $ outputs the real and imaginary parts of beamforming, which should contain both positive and negative values. Thus, $\text{AL}\left(\cdot \right) $ in $\mathcal{C}\left(\cdot,\theta_L \right) $  can adopt  $\text{Tanh}(x)=\frac{e^{x}-e^{-x}}{e^{x}+e^{-x}}$.

\emph{Remark 3: For the constraint $ \left(Q_t,I_t\right)\neq \left(Q_{t^\prime}, I_{t^\prime} \right), \forall (t\neq t^\prime)\in\mathcal{T} $ in the optimization problem (\ref{eq3}),  when the dimension of the input 3D real-valued CSI $\mathbf{H}_{t,\text{mod}}^{\text{3D}}$ at the $t^{th} $ period is $Q_t \times I_t \times MN$, the dimension of the corresponding output beamforming should be a 3D complex-valued tensor of dimension $Q_t\times I_t \times M$, which can be transformed into a 3D real-valued tensor of dimension $Q_t\times I_t \times 2M$. At the $ {t^\prime}^{th} $ period with $  \forall (t\neq t^\prime)\in\mathcal{T} $, the dimension of the input 3D real-valued CSI $\mathbf{H}_{t^\prime,\text{mod}}^{\text{3D}}$ changes from $Q_t \times I_t \times MN$ to $Q_{t^\prime} \times I_{t^\prime} \times MN$, then the dimension of the corresponding output beamforming  should vary from $Q_t \times I_t \times 2M$ to $Q_{t^\prime} \times I_{t^\prime} \times 2M$.}

Based on \emph{Remark 3}, when  $\mathbf{H}_{t,\text{mod}}^{\text{3D}}\in \mathbb{R}^{Q_t \times I_t \times MN}$ is inputted to $\mathcal{C}\left(\cdot, \theta_l\right),l=\{1,\cdots,L\} $, the dimension of the output $ \mathbf{C}_L $ of the $ L^{th} $ layer $\mathcal{C}\left(\cdot, \theta_L\right)$ should be $Q_t\times I_t \times 2M$. However, the dimension of $ \mathbf{C}_L $  is determined by the architectural parameters of the CL in  $\mathcal{C}\left(\cdot, \theta_l\right),l=\{1,\cdots,L\} $ such as the number of convolution kernel and the sizes of convolution kernel, sliding step, zero padding. Consequently, in what follows, we derive some architectural conditions for the CL  in  $\mathcal{C}\left(\cdot, \theta_l\right),l=\{1,\cdots,L\} $ to satisfy \emph{Remark 3}.

\emph{Proposition 1: Let $w_{\mathcal{C}\left(\cdot,\theta_l \right)}^{\text{in}}$,  $h_{\mathcal{C}\left(\cdot,\theta_l \right)}^{\text{in}}$, $w_{\mathcal{C}\left(\cdot,\theta_l \right)}^{\text{out}}$ and $h_{\mathcal{C}\left(\cdot,\theta_l \right)}^{\text{out}}$ denote the input-output width and height of $\mathcal{C}\left(\cdot,\theta_l \right) $, as well as $k_l^w$, $k_l^h$, $p_l^w$, $p_l^h$, $s_l^w$ and $s_l^h$ denote the width and height of the convolution kernel, the zero padding, the sliding step for the CL of $\mathcal{C}\left(\cdot,\theta_l \right) $, respectively. When $s_l^w=1$, $s_l^h=1$, if $p_l^w=\frac{1}{2}(k_l^w-1)$, $p_l^h=\frac{1}{2}(k_l^h-1)$, both $p_l^w$,  $p_l^h$ and $k_l^w$, $k_l^h$ are positive integers, then $w_{\mathcal{C}\left(\cdot,\theta_l \right)}^{\text{out}}=w_{\mathcal{C}\left(\cdot,\theta_l \right)}^{\text{in}}$ and  $h_{\mathcal{C}\left(\cdot,\theta_l \right)}^{\text{out}}=h_{\mathcal{C}\left(\cdot,\theta_l \right)}^{\text{in}}$.}

\emph{Proof:} As can be seen in Fig.1, $\mathcal{C}\left(\cdot,\theta_l \right) $ includes one CL, BN and AL. For the CL in $\mathcal{C}\left(\cdot,\theta_l \right) $, its output width and height $w_{\mathcal{C}\left(\cdot,\theta_l \right)}^{\text{CL}} \times h_{\mathcal{C}\left(\cdot,\theta_l \right)}^{\text{CL}} $ are denoted as
\begin{equation}
	\begin{cases}
		w_{\mathcal{C}\left(\cdot,\theta_l \right)}^{\text{CL}}=\frac{w_{\mathcal{C}\left(\cdot,\theta_l \right)}^{\text{in}}+2 p_{l}^w-k_{l}^w}{s_l^w}+1,\\ 
		h_{\mathcal{C}\left(\cdot,\theta_l \right)}^{\text{CL}}\thinspace=\frac{h_{\mathcal{C}\left(\cdot,\theta_l \right)}^{\text{in}}\thinspace+2 p_l^h-k_{l}^h\thinspace}{s_l^h}+1,
		\label{eq31}
	\end{cases}
\end{equation}
where $s_l^w=1$, $s_l^h=1$, $p_l^w=\frac{1}{2}(k_l^w-1)$ and $p_l^h=\frac{1}{2}(k_l^h-1)$ are brought into Eq.(\ref{eq31}), 
\begin{equation}
	\begin{cases}
		w_{\mathcal{C}\left(\cdot,\theta_l \right)}^{\text{CL}}=\frac{w_{\mathcal{C}\left(\cdot,\theta_l \right)}^{\text{in}}+2\times \frac{1}{2}(k_l^w-1)-k_{l}^w}{1}+1=w_{\mathcal{C}\left(\cdot,\theta_l \right)}^{\text{in}},\\ 
		h_{\mathcal{C}\left(\cdot,\theta_l \right)}^{\text{CL}}\thinspace=\frac{h_{\mathcal{C}\left(\cdot,\theta_l \right)}^{\text{in}}\thinspace+2\times \frac{1}{2}(k_l^h-1)-k_{l}^h\thinspace}{1}+1=h_{\mathcal{C}\left(\cdot,\theta_l \right)}^{\text{in}}.
		\label{eq32}
	\end{cases}
\end{equation}
Based on Eq.(\ref{eq32}), the output width and height of the CL in $\mathcal{C}\left(\cdot,\theta_l \right) $ are $w_{\mathcal{C}\left(\cdot,\theta_l \right)}^{\text{in}}$  and $h_{\mathcal{C}\left(\cdot,\theta_l \right)}^{\text{in}}$, respectively. On the other hand, the BN and AL do not change the input dimension, i.e., the output width and height of the BN and AL in  $\mathcal{C}\left(\cdot,\theta_l \right) $ are also  $w_{\mathcal{C}\left(\cdot,\theta_l \right)}^{\text{in}}$  and $h_{\mathcal{C}\left(\cdot,\theta_l \right)}^{\text{in}}$, respectively. Consequently, the output width and height of $\mathcal{C}\left(\cdot,\theta_l \right) $ are  $w_{\mathcal{C}\left(\cdot,\theta_l \right)}^{\text{in}}$  and $h_{\mathcal{C}\left(\cdot,\theta_l \right)}^{\text{in}}$,  i.e., $w_{\mathcal{C}\left(\cdot,\theta_l \right)}^{\text{out}}=w_{\mathcal{C}\left(\cdot,\theta_l \right)}^{\text{in}}$ and  $h_{\mathcal{C}\left(\cdot,\theta_l \right)}^{\text{out}}=h_{\mathcal{C}\left(\cdot,\theta_l \right)}^{\text{in}}$, respectively. $\hfill\blacksquare$

\emph{Proposition 2: When $s_l^w>1$, $s_l^h>1$, if $p_l^w=\frac{1}{2}(w_{\mathcal{C}\left(\cdot,\theta_l \right)}^{\text{in}}s_l^w-w_{\mathcal{C}\left(\cdot,\theta_l \right)}^{\text{in}}-s_l^w+k_l^w)$ and $p_l^h=\frac{1}{2}(h_{\mathcal{C}\left(\cdot,\theta_l \right)}^{\text{in}}s_l^h-h_{\mathcal{C}\left(\cdot,\theta_l \right)}^{\text{in}}-s_l^h+k_l^h)$, as well as $p_l^w$,  $p_l^h$, $s_l^w$, $s_l^h$, $k_l^w$, $k_l^h$ are positive integers, then $w_{\mathcal{C}\left(\cdot,\theta_l \right)}^{\text{out}}=w_{\mathcal{C}\left(\cdot,\theta_l \right)}^{\text{in}}$ and  $h_{\mathcal{C}\left(\cdot,\theta_l \right)}^{\text{out}}=h_{\mathcal{C}\left(\cdot,\theta_l \right)}^{\text{in}}$.}

\emph{Proof:} For the CL in $\mathcal{C}\left(\cdot,\theta_l \right) $, where $s_l^w>1$, $s_l^h>1$, $p_l^w=\frac{1}{2}(w_{\mathcal{C}\left(\cdot,\theta_l \right)}^{\text{in}}s_l^w-w_{\mathcal{C}\left(\cdot,\theta_l \right)}^{\text{in}}-s_l^w+k_l^w)$ and $p_l^h=\frac{1}{2}(h_{\mathcal{C}\left(\cdot,\theta_l \right)}^{\text{in}}s_l^h-h_{\mathcal{C}\left(\cdot,\theta_l \right)}^{\text{in}}-s_l^h+k_l^h)$, its output width and height  $w_{\mathcal{C}\left(\cdot,\theta_l \right)}^{\text{CL}} \times h_{\mathcal{C}\left(\cdot,\theta_l \right)}^{\text{CL}} $ are denoted as
\begin{equation}
	\begin{cases}
		w_{\mathcal{C}\left(\cdot,\theta_l \right)}^{\text{CL}}=& \frac{w_{\mathcal{C}\left(\cdot,\theta_l \right)}^{\text{in}}+2\times \frac{1}{2}(w_{\mathcal{C}\left(\cdot,\theta_l \right)}^{\text{in}}s_l^w-w_{\mathcal{C}\left(\cdot,\theta_l \right)}^{\text{in}}-s_l^w+k_l^w)-k_{l}^w}{s_l^w}\\&+1
		=w_{\mathcal{C}\left(\cdot,\theta_l \right)}^{\text{in}},\\
		h_{\mathcal{C}\left(\cdot,\theta_l \right)}^{\text{CL}}=& \frac{h_{\mathcal{C}\left(\cdot,\theta_l \right)}^{\text{in}}+\thinspace2\times \frac{1}{2}(h_{\mathcal{C}\left(\cdot,\theta_l \right)}^{\text{in}}s_l^h-\thinspace h_{\mathcal{C}\left(\cdot,\theta_l \right)}^{\text{in}}-\thinspace s_l^h+\thinspace k_l^h)-\thinspace k_{l}^h}{s_l^h}\\&+1
		=h_{\mathcal{C}\left(\cdot,\theta_l \right)}^{\text{in}}.
		\label{eq33}
	\end{cases}
\end{equation}
Similarly, the output width and height of $\mathcal{C}\left(\cdot,\theta_l \right) $ are  $w_{\mathcal{C}\left(\cdot,\theta_l \right)}^{\text{in}}$  and $h_{\mathcal{C}\left(\cdot,\theta_l \right)}^{\text{in}}$,  i.e., $w_{\mathcal{C}\left(\cdot,\theta_l \right)}^{\text{out}}=w_{\mathcal{C}\left(\cdot,\theta_l \right)}^{\text{in}}$ and  $h_{\mathcal{C}\left(\cdot,\theta_l \right)}^{\text{out}}=h_{\mathcal{C}\left(\cdot,\theta_l \right)}^{\text{in}}$. $\hfill\blacksquare$

Obviously, as long as the architectures $k_l^w$, $k_l^h$, $p_l^w$, $p_l^h$, $s_l^w$ and $s_l^h$ in each $\mathcal{C}\left(\cdot, \theta_l\right)$ satisfy \emph{Proposition 1} or \emph{Proposition 2}, the width and height of  $ \mathbf{C}_L $ are $ Q_t \times I_t $. In addition, if the number of convolutional kernels $ c_L $ for the $ L^{th} $ layer $\mathcal{C}\left(\cdot, \theta_L\right)$  is set to $ 2M $, then the dimension of  $ \mathbf{C}_L $ is $ Q_t \times I_t \times 2M $. On the other hand, based on \emph{Proposition 1} or \emph{Proposition 2}, when the dimension of the input 3D real-valued CSI $\mathbf{H}_{t^\prime,\text{mod}}^{\text{3D}}$  changes from $Q_t \times I_t \times MN$ to $Q_{t^\prime} \times I_{t^\prime} \times MN$ at the $ t^\prime $ period with $  \forall (t\neq t^\prime)\in\mathcal{T} $, then the dimension of  $ \mathbf{C}_L $ also varies  from $Q_t \times I_t \times 2M$ to $Q_{t^\prime} \times I_{t^\prime} \times 2M$. Consequently, the architectural conditions that satisfy  the constraint $ \left(Q_t,I_t\right)\neq \left(Q_{t^\prime}, I_{t^\prime} \right), \forall (t\neq t^\prime)\in\mathcal{T} $ in the optimization problem (\ref{eq3}) are summarized in \emph{Remark 4}.

\emph{Remark 4: The constraint $ \left(Q_t,I_t\right)\neq \left(Q_{t^\prime}, I_{t^\prime} \right), \forall (t\neq t^\prime)\in\mathcal{T} $ in the optimization problem (\ref{eq3}) is satisfied, if only the following conditions  are held as
	\begin{itemize}
		\item[1)] The architectural parameters $k_l^w$, $k_l^h$,  $p_l^w$, $p_l^h$, $s_l^w$, $s_l^h$, $l=1,\cdots, L$ in each $\mathcal{C}\left(\cdot,\theta_l \right) $ satisfy Proposition 1 or Proposition 2.
		\item[2)] The number of convolutional kernels $c_L$ for  $\mathcal{C}\left(\cdot,\theta_L \right) $ is equal to $2M$.
\end{itemize}}

In summary, based on \emph{Proposition 1}, \emph{Proposition 2} and \emph{Remark 4}, as long as the two conditions in \emph{Remark 4}  are satisfied,  $\mathcal{C}\left(\cdot, \theta_l\right)$ maps  CSI  to beamforming with satisfying $ \left(Q_t,I_t\right)\neq \left(Q_{t^\prime}, I_{t^\prime} \right), \forall (t\neq t^\prime)\in\mathcal{T} $ in the optimization problem (\ref{eq3}).

\subsection{High-Generalization Beamforming Module $\mathcal{G}\left(\cdot, \theta_l\right)$}

$\mathcal{C}\left(\cdot, \theta_l\right)$ have been solved for dynamic wireless environments with the varying numbers of APs and users $ \left(Q_t,I_t\right)\neq \left(Q_{t^\prime}, I_{t^\prime} \right), \forall (t\neq t^\prime)\in\mathcal{T} $  in cell-free systems as long as the two conditions in \emph{Remark 4} are satisfied. In the following,  under the conditions specified in \emph{Remark 4}, $\mathcal{G}\left(\cdot, \theta_l\right)$ is designed to improve the generalization performance of $ D\left(\mathbf{H}_t \right) \neq D\left(\mathbf{H}_{t^\prime} \right), \forall (t\neq t^\prime)\in\mathcal{T} $ in the optimization problem (\ref{eq3}), by processing the output $\mathbf{C}_{l}$  of $\mathcal{C}\left(\cdot, \theta_l\right)$  to obtain the valuable features of the varying channels. Especially,  $\mathcal{G}\left(\cdot, \theta_l\right)$ inclueds a distinguishing sensitive feature module and a discarding sensitive feature module.

\subsubsection{Distinguishing Sensitive Feature Module} 
 $\mathcal{C}\left(\cdot, \theta_l\right)$ aims at the real-time beamforming  design, whereas the purpose of $\mathcal{G}\left(\cdot, \theta_l\right)$ is to obtain the valuable features of the varying channels. These are two different tasks. To address this, a commonly used simple gradient reversal layer (GRL) \cite{ref6} is added between these two tasks.  The GRL transforms the gradient into a negative gradient in the gradient backpropagation, which  forms an adversarial training  to find a balance between the above two tasks, and please refer to \cite{ref6} for more details.

On the other hand,  the output $\mathbf{C}_{l}\in \mathbb{R}^{Q_t \times I_t \times C}$ \footnote{Note that \emph{Remark 4} only requires the third dimension of $ \mathbf{C}_{L} $ to be $ 2M $, and has no requirement for the other $ \mathbf{C}_{l},l=\{1,\cdots,L-1\} $. For simplicity,  let $ C $ be collectively referred to the third dimension of $ \mathbf{C}_{l} $, i.e., $\mathbf{C}_{l}\in \mathbb{R}^{Q_t \times I_t \times C}$.}  of  $\mathcal{C}\left(\cdot, \theta_l\right)$ contains $ C $ features with dimension $ {Q_t \times I_t} $. To determine which  features contain sensitive information for the varying channels, this work designs a sensitive feature discriminator $\mathcal{D}\left(\cdot, \theta_l\right)$ containing a global average pooling (GAP) layer and a fully connected (FC) layer. Specifically,  $\mathcal{D}\left(\cdot, \theta_l\right)$  first utilizes GAP to process  $\mathbf{C}_{l}\in \mathbb{R}^{Q_t \times I_t \times C}$ to obtain global features $ \mathbf{g}_{l} $ for reducing computational complexity, which is denoted as 
\begin{equation}
	\mathbf{g}_{l}=\text{GAP}\left(\mathbf{C}_{l}\right)\in \mathbb{R}^{C}.
	\label{eq5}
\end{equation}
Afterwards, $ \mathbf{g}_{l} $ is fed into the FC layer,  where the number of the input-output neurons of the FC layer are $ C $ and $ T $, respectively. Meanwhile, a sensitive feature discriminator loss function $ \mathcal{L}_l^{\text{dis}} $ is minimized to train $\mathcal{D}\left(\cdot, \theta_l\right)$, i.e., 
\begin{equation}
	\mathcal{L}_l^{\text{dis}}=-\frac{1}{T}\sum_{t=1}^{T}\sum_{t^\prime=1}^{T}{\mathbbm{1}}_{[t=t^\prime]} \log \mathcal{D}\left(\mathbf{g}_{l},\theta_l\right),
	\label{eq6}
\end{equation}
where $ {\mathbbm{1}}_{[t=t^\prime]} $  denotes an indicator function, i.e., if the index $ t $ is equal to $ t^\prime $, then $ {\mathbbm{1}}_{[t=t^\prime]}=1 $, otherwise  $ {\mathbbm{1}}_{[t=t^\prime]} =0 $. 

Note that an intuitive finding can be obtained for $\mathcal{D}\left(\cdot, \theta_l\right)$. That is, the features contributing most to the prediction of $\mathcal{D}\left(\cdot, \theta_l\right)$ may contain sensitive features to the varying channels, since the inputs and outputs of $\mathcal{D}\left(\cdot, \theta_l\right)$ are the features and the number of the varying channels, respectively. Consequently, this paper defines a score $ \mathbf{s}_l $ to represent the contribution of the features, computed as the dot product of $ \mathbf{g}_{l}\in \mathbb{R}^{C} $ and the weighted activations $ \mathbf{w}_l^{\text{cor}}\in \mathbb{R}^{C} $ of the FC layer for the correct prediction of $\mathcal{D}\left(\cdot, \theta_l\right)$, as the weighted activations indicate the importance of the input data \cite{ref13}.  Formally, it can be denoted as
\begin{equation}
	\mathbf{s}_l=\mathbf{w}_l^{\text{cor}}\otimes\mathbf{g}_{l}\in\mathbb{R}^{C},
	\label{eq7}
\end{equation}
where $ \otimes $ denotes the dot product. Obviously, when the value of the element in $ \mathbf{s}_l \in \mathbb{R}^{C} $ is larger, its corresponding feature is more sensitive to the varying channels, and vice versa.

\subsubsection{Discarding Sensitive Feature Module}
Based on $ \mathbf{s}_l \in \mathbb{R}^{C} $, this work explicitly discards some sensitive features to the varying channels during the training stage. Specifically, given  $ \mathbf{s}_l \in \mathbb{R}^{C} $, this work computes the probability $ p_l^c $ of discarding the $ c^{th} $ feature of dimension $ Q_t \times I_t$ in  $\mathbf{C}_{l}\in \mathbb{R}^{Q_t \times I_t \times C}$, i.e., 
\begin{equation}
	p_l^c=\frac{s_l^c}{\sum\limits_{c=1}^C{s_l^c}},
	\label{eq8}
\end{equation} 
where $ s_l^c $ denotes the $ c^{th} $ element in  $ \mathbf{s}_l= \left[s_l^1,\cdots, s_l^c,\cdots, s_l^C  \right]\in \mathbb{R}^{C} $. 

Subsequently, based on $ \mathbf{p}_l=\left[p_l^1,\cdots, p_l^c,\cdots, p_l^C  \right] \in \mathbb{R}^{C}$, a weighted random selection (WRS) algorithm \cite{ref15} is applied  to generate the binary mask, since it is highly computationally efficient with complexity $ O(C) $. To be more concrete, for the $ c^{th} $ feature of dimension $ Q_t \times I_t$ in  $\mathbf{C}_{l}\in \mathbb{R}^{Q_t \times I_t \times C}$ with probability $ p_l^c $, a random number $ r_l^c\in\left(0,1 \right)  $ is generated, where a key value $ k_l^c $ is computed as 
\begin{equation}
	k_l^c= {r_l^c}^{\frac{1}{p_l^c}}. 
	\label{eq9}
\end{equation} 
When  $ s_l^c $  is larger, both $ p_l^c $  and  $ k_l^c $  increase, indicating that the feature of dimension $ Q_t \times I_t$ in  $\mathbf{C}_{l}\in \mathbb{R}^{Q_t \times I_t \times C}$ is more sensitive, and vice versa. Consequently,  to discard the sensitive features and retain the  valuable features for the varying channels, the  $ C_{\text{dis}} $ items with the largest key
values are selected, and their corresponding mask values are set to 0, i.e.,
\begin{equation}
	m_l^c= \begin{cases}0 & \text {if} \quad c\in{\text{TOP}}\left(\left[k_l^1,\cdots, k_l^c,\cdots, k_l^C  \right], C_{\text{dis}} \right),\\ 1 & \text {otherwise, }\end{cases}
	\label{eq10}
\end{equation}
where $ c\in{\text{TOP}}\left(\left[k_l^1,\cdots, k_l^c,\cdots, k_l^C  \right], C_{\text{dis}} \right) $ denotes the $ C_{\text{dis}} $ items with the largest key.

Afterwards, the binary mask $ \mathbf{m}_l= \left[m_l^1,\cdots, m_l^c,\cdots, m_l^C  \right]\in \mathbb{R}^{C} $ is dot-multiplied with $\mathbf{C}_{l}$ to obtain the valuable features  $ \mathbf{G}_{l} $ for the varying channels, which is defined as
\begin{equation}
	\mathbf{G}_{l}= \mathbf{m}_l \otimes \mathbf{C}_{l}\in \mathbb{R}^{Q_t \times I_t \times C}.
	\label{eq11}
\end{equation} 
It is obvious that $ \mathbf{G}_{l}$  effectively  extracts the valuable features for the varying channels via the binary mask to discard the sensitive features. Remarkably, $\mathcal{G}\left(\cdot, \theta_l\right)$ is only added after $ \mathbf{C}_l, l=\{1,\cdots,L-1\} $ without  $ \mathbf{C}_L $ to  guarantee that the output beamforming  of  HGNet is a real-valued tensor of dimension  $Q_t \times I_t \times 2M $ with $\mathbf{H}_{t,\text{mod}}^{\text{3D}}\in \mathbb{R}^{Q_t \times I_t \times MN}$ as the input.

\subsection{Output}
Before the output  beamforming,  an identity mapping  is added to construct the residual structure  for effectively avoiding the gradient disappearance problem \cite{ref7},  which is denoted as
\begin{equation}
	\mathbf{V}_{t,\text{rea}}^{\text{3D}}=\text{AL}\left(\mathbf{C}_L+\mathbf{V}_{\text{IM}} \right)\in \mathbb{R}^{Q_t \times I_t \times 2M},
	\label{eq12}
\end{equation} 
where $\mathbf{V}_{\text{IM}}$  denotes the output of the  identity  mapping with  $\mathbf{H}_{t,\text{mod}}^{\text{3D}}\in \mathbb{R}^{Q_t \times I_t \times MN}$ as the input.  Afterwards, the output module first transforms $\mathbf{V}_{t,\text{rea}}^{\text{3D}}\in \mathbb{R}^{Q_t \times I_t \times 2M}$ into a 3D complex-valued beamforming tensor $ \mathbf{V}_{t,\text{com}}^{\text{3D}}\in \mathbb{C}^{Q_t \times I_t \times M} $ as follows
\begin{equation}
	\mathbf{V}_{t,\text{com}}^{\text{3D}}=\mathbf{V}_{t,\text{rea}}^{\text{3D}}[:,:,0:M]+j\mathbf{V}_{t,\text{rea}}^{\text{3D}}[:,:,M:2M].
	\label{eq13}
\end{equation}
On the other hand, $ \mathbf{V}_{t,\text{com}}^{\text{3D}} $ also needs to satisfy the power constraint. Since this is a convex constraint \cite{ref8}, it can be satisfied using a projection function. Consequently, following \cite{ref8}, the output module applies the following projection function to satisfy the power constraint, i.e., 
\begin{equation}
	\mathbf{v}_{i,t}^q= \begin{cases}\mathbf{v}_{i,t}^q & \text { if } \sum\limits_{i=1}^{I_t}\mathbf{v}_{i,t}^{q,H}\mathbf{v}_{i,t}^{q}\leq \text{P}_{\textrm{max}},\\ \frac{\mathbf{v}_{i,t}^q}{\sum\limits_{i=1}^{I_t}\mathbf{v}_{{i,t}}^{q,H}\mathbf{v}_{{i,t}}^{q}} {\text{P}_{\text{max}}} & \text { otherwise.}\end{cases}
	\label{eq14}
\end{equation}
Finally, following the commonly utilized unsupervised training method \cite{bib34}, we also take  the negative of the sum rate as the loss function to train HGNet, where the output beamforming of HGNet is denoted as $\mathbf{V}_{t,\text{HGNet}}^{\text{3D}}\in \mathbb{C}^{Q_t \times I_t \times M}$.

\subsection{Theoretical Analysis of High-Generalization of HGNet}

In this subsection, the bias of the varying channels is measured by the commonly known maximum mean discrepancy (MMD) distance. It is a metric that quantifies the difference between the data distribution in the source and the target domains\cite{bib35}. It is defined as follows.

\emph{Definition 1: (MMD \cite{bib24}) Let $ \mathcal{F}=\left\{ f\in\mathcal{H_{\mathrm{\mathit{k}}}}:\left\Vert f\right\Vert _{H_{\mathrm{\mathit{k}}}}\leq1\right\}  $ denote the set of functions on the sample space, in which $ \mathcal{H_{\mathrm{\mathit{k}}}} $ is a reproducing kernel Hilbert space (RKHS) with kernel function $ k $. Let  $ \mathcal{S}_{\text{tr}} $ and $ \mathcal{T}_{\text{in}} $ denote the source domain of the training stage and the target domain  of the inference stage, where $ \hat{\mathcal{S}}_{\text{tr}}$ and $\hat{\mathcal{T}}_{\text{in}} $ denote the $ D $ data sampled from $ \mathcal{S}_{\text{tr}} $ and $ \mathcal{T}_{\text{in}} $, respectively. The MMD distance between  $ \hat{\mathcal{S}}_{\text{tr}}$ and $\hat{\mathcal{T}}_{\text{in}} $ is defined as
$$d_{\text{MMD}}\left(\hat{\mathcal{S}}_{\text{tr}},\hat{\mathcal{T}_{\text{in}}}\right)	=\underset{f\in\mathcal{F}}{\textrm{sup}}\left[\frac{1}{D}\sum_{d=1}^{D}f\left(x_{d}\right)-\frac{1}{D}\sum_{d=1}^{D}f\left(y_{d}\right)\right],$$
where $ x_d $ and $ y_d $ denote the $ d^{th} $ data in $ \hat{\mathcal{S}}_{\text{tr}}$ and $\hat{\mathcal{T}}_{\text{in}}$, respectively.}

To facilitate analysis for  the high generalization  of $\mathcal{G}\left(\cdot, \theta_l\right)$ in HGNet, similar to MMD in \emph{Definition 1}, a G-MMD is defined as follows.

\emph{Definition 2: (G-MMD) Following the definition in MMD,  G-MMD distance between  $ \hat{\mathcal{S}}_{\text{tr}}$ and $\hat{\mathcal{T}}_{\text{in}} $ is defined as
	\begin{align*}
		&d_{\text{G-MMD}}\left(\hat{\mathcal{S}}_{\text{tr}},\hat{\mathcal{T}}_{\text{in}}\right)=\\& \frac{1}{C}\sum_{c=1}^{C}\underset{f_{c}\in\mathcal{F_{\mathit{c}}}}{\textrm{sup}}\left[\frac{1}{D}\sum_{d=1}^{D}f_c\left(\textbf{G}_{\mathit{l}}[:,:,c]\right)_{\hat{\mathcal{S}}_{\text{tr}}}^d-\frac{1}{D}\sum_{d=1}^{D}f_c\left(\textbf{G}_{\mathit{l}}[:,:,c]\right)_{\hat{\mathcal{T}}_{\text{in}}}^d\right],	
	\end{align*}
where $ \left(\textbf{G}_{\mathit{l}}[:,:,c]\right)_{\hat{\mathcal{S}}_{\text{tr}}}^d $ and $ \left(\textbf{G}_{\mathit{l}}[:,:,c]\right)_{\hat{\mathcal{T}}_{\text{in}}}^d $ denote the $ c^{th} $ features of dimension $ Q_t \times I_t $ in the output of $\mathcal{G}\left(\cdot, \theta_l\right)$ when the $ d^{th} $ data in $ \hat{\mathcal{S}}_{\text{tr}}$ and $\hat{\mathcal{T}}_{\text{in}} $ are inputted to  HGNet, respectively. $ \mathcal{F}_c=\left\{ f_c\in\mathcal{H_{\mathrm{\mathit{k}}}}:\left\Vert f_c\right\Vert _{H_{\mathrm{\mathit{k}}}}\leq1\right\}  $ denotes the set of functions on the sample space corresponding to $ \textbf{G}_{\mathit{l}}[:,:,c] $.}

\emph{Proposition 3: With the previous
	definitions,  let  $ \mathcal{S}_{\text{tr}}^t , t\in \mathcal{T}$ denote the source domain of the training stage for $ T $ varying channels, where $ \hat{\mathcal{S}}_{\text{tr}}^t$ denote the $ D $ data sampled from $ \mathcal{S}^t $.	Let  $ R_{\mathcal{T}_{\text{in}}}\left[\mathcal{G}\right] $ denote the generalized risk bound of  $\mathcal{G}\left(\cdot, \theta_l\right)$ in HGNet for the target domain $ {\mathcal{T}}_{\text{in}}$. With the probability of at least $ 1 -\delta, \delta \in (0,1) $, the following inequality holds for $ R_{\mathcal{T}_{\text{in}}}\left[\mathcal{G}\right] $: 
	\begin{align*}
	R_{\mathcal{T}_{\text{in}}}\left[\mathcal{G}\right]\leq&\underset{t,t^{\prime}\in\mathcal{T}}{\textrm{sup}}d_{\text{G-MMD}}\left(\mathcal{S}_{\text{tr}}^{\mathit{t^{\prime}}},\mathcal{S}_{\text{tr}}^{\mathit{t}}\right)+d_{\text{G-MMD}}\left(\hat{\overline{\mathcal{T}_{\text{in}}}},\overline{\mathcal{T}_{\text{in}}}\right)\\&+\varpi+\varrho+\xi,	
\end{align*}
where $ \overline{\mathcal{T}_{\text{in}}}=\sum_{t=1}^{T}\kappa_{t}\mathcal{S}_{\text{tr}}^t$ denotes a mixture of $ T $ varying channels closest to the target domain $ \mathcal{T}_{\text{in}} $, and the mixture weight is given by $ \sum_{t=1}^{T} \kappa_{t}=1 $. $ \hat{\overline{\mathcal{T}_{\text{in}}}} $ denotes the $ D $ data sampled from $ \overline{\mathcal{T}_{\text{in}}} $. $ \varpi= \sum_{t=1}^{T}\kappa_{t}R_{{\mathcal{S}}_{\text{tr}}^t}\left[{\mathcal{G}}\right] $  denotes the mixture weight of the generalization error for known source domains  $ \mathcal{S}_{\text{tr}}^t , t\in\mathcal{T}$. $ \varrho= R_{\overline{\mathcal{T}_{\text{in}}}}\left[\mathcal{G^{*}}\right]+R_{\mathcal{T}_{\text{in}}}\left[\mathcal{G^{*}}\right] $ denotes the combined error of ideal $\mathcal{G}^*\left(\cdot, \theta_l^*\right)$. $ \xi=\frac{2}{D}\left(\sum_{t=1}^{T}\kappa_{t}{\mathbb{E}}\left[\sqrt{\text{tr}\left(k_{{\hat{\mathcal{S}}}_{\text{tr}}^{t}}\right)}\right]+{\mathbb{E}}\left[\sqrt{\text{tr}\left(k_{\hat{\mathcal{T}_{\text{in}}}}\right)}\right]\right)+2\sqrt{\frac{\log\left(\frac{2}{\delta}\right)}{2D}} $, in which  $ k_{{\hat{\mathcal{S}}}_{\text{tr}}^{t}} $ and $ k_{\hat{\mathcal{T}_{\text{in}}}} $ denote kernel functions computed on samples from  $ \hat{\mathcal{S}}_{\text{tr}}^t$ and  $\hat{\mathcal{T}_\text{in}} $, respectively.}

\emph{Proof: Please see Appendix A for the detailed proof.} $\hfill\blacksquare$

From \emph{Proposition 3}, it is clear that $\underset{t,t^{\prime}\in\mathcal{T}}{\textrm{sup}}d_{\text{G-MMD}}\left(\mathcal{S}_{\text{tr}}^{\mathit{t^{\prime}}},\mathcal{S}_{\text{tr}}^{\mathit{t}}\right)$  and  $d_{\text{G-MMD}}\left(\hat{\overline{\mathcal{T}_{\text{in}}}},\overline{\mathcal{T}_{\text{in}}}\right) $ mainly determine the upper bound  of the generalization error of $\mathcal{G}\left(\cdot, \theta_l\right)$ in HGNet.  To be
 concrete, $\underset{t,t^{\prime}\in\mathcal{T}}{\textrm{sup}}d_{\text{G-MMD}}\left(\mathcal{S}_{\text{tr}}^{\mathit{t^{\prime}}},\mathcal{S}_{\text{tr}}^{\mathit{t}}\right)$ denotes the G-MMD distance of the output $ \mathbf{G}_l $ of $\mathcal{G}\left(\cdot, \theta_l\right)$ for any pair of the varying channels in the source domain of the training stage. Since the output $ \mathbf{G}_l $ of $\mathcal{G}\left(\cdot, \theta_l\right)$  explicitly discards the sensitive features to the varying channels  during the training stage, this can promote $\mathcal{G}\left(\cdot, \theta_l\right)$ to learn a model that extracts non-sensitive features for the varying channels, i.e., effectively reducing $\underset{t,t^{\prime}\in\mathcal{T}}{\textrm{sup}}d_{\text{G-MMD}}\left(\mathcal{S}_{\text{tr}}^{\mathit{t^{\prime}}},\mathcal{S}_{\text{tr}}^{\mathit{t}}\right)$. On the other hand, $d_{\text{G-MMD}}\left(\hat{\overline{\mathcal{T}_{\text{in}}}},\overline{\mathcal{T}_{\text{in}}}\right) $ denotes  the G-MMD distance of the output $ \mathbf{G}_l $ of $\mathcal{G}\left(\cdot, \theta_l\right)$ for source domain  $ \mathcal{S}_{\text{tr}}^t , t\in \mathcal{T}$ and target domain $ \mathcal{T}_{\text{in}} $. After removing the sensitive features to the varying channels, the features extracted from the target domain $ \mathcal{T}_{\text{in}} $ would become more similar to those of the source domains $ \mathcal{S}_{\text{tr}}^t , t\in \mathcal{T}$, thus also decreasing $d_{\text{G-MMD}}\left(\hat{\overline{\mathcal{T}}},\overline{\mathcal{T}}\right) $. In summary, based on \emph{Proposition 3}, $\mathcal{G}\left(\cdot, \theta_l\right)$  obtains a lower upper bound of  the generalization error, which guarantees that HGNet yields a better generalization sum rate performance for $ D\left(\mathbf{H}_t \right) \neq D\left(\mathbf{H}_{t^\prime} \right), \forall (t\neq t^\prime)\in\mathcal{T} $ in the optimization problem (\ref{eq3}).

\section{OAU Algorithm}
To realize the online adaptive real-time beamforming design to further improve the sum rate performance of dynamic wireless environments with the varying channels and the different numbers of APs and users in cell-free systems, the OAU algorithm is proposed in this section. Intuitively, it is a natural choice to  adaptively update the parameters of HGNet online by taking the negative of the sum rate as a loss function.  Despite the simplicity of this approach, this suffers from two major problems. 
\begin{itemize}
	\item[1)]  The parameters  of HGNet are normally high-dimensional parameters, and updating the entire parameters is time-consuming without fulfilling the requirements of the real-time beamforming design.	
\end{itemize}

\begin{itemize}
	\item[2)] 
	Since computing the sum rate involves the matrix inversion with high computational complexity, it is also difficult to satisfy the real-time beamforming design when updating the parameters of HGNet with the negative sum rate as the loss function.
\end{itemize}

To solve the first problem, the proposed  OAU algorithm online  adaptively updates the affine parameters of  the BN layer in HGNet instead of the entire parameters of HGNet.  This is because  the affine parameters comprise  less than 3\% of the total number of  parameters in  HGNet. Consequently, updating the affine parameters of the BN layer is more computationally efficient, and suitable for the real-time beamforming design.  Concretely, to simplify notation, the input of the BN layer at the  $l^{th}$ layer $\mathcal{C}\left(\cdot, \theta_l\right)$  of HGNet  is defined as $ \mathbf{X}_{l}=\text{CL}\left(\mathbf{G}_{l-1}, \theta_l\right)\in \mathbb{R}^{Q_t \times I_t \times C} $. The BN layer first calculates the normalized value of $ \mathbf{X}_{l}\in \mathbb{R}^{Q_t \times I_t \times C} $ as
\begin{equation}\label{eq15}
\mathbf{X}_{l}^{\text{nor}}=\frac{  \mathbf{X}_{l}-\text{E}[\mathbf{X}_{l}]}{\sqrt{\text{Var}[\mathbf{X}_{l}]+\epsilon}}, 
\end{equation}
where  $ \epsilon $ denotes a pretty small constant to avoid a zero in the denominator. $ \text{E}[\mathbf{X}_{l}] $ and $ \text{Var}[\mathbf{X}_{l}] $  denote the mean and variance of $ \mathbf{X}_{l} $, respectively. Since the input CSI in the inference stage is a single input rather than a mini-batch as in the training phase, it is more challenging to compute $\text{E}[\mathbf{X}_{l}] $ and $ \text{Var}[\mathbf{X}_{l}] $. As a result, following the widely applied approach of deep learning, the unbiased estimation for the mean and variance of the mini-batch of the training phase over the total training dataset is used as $\text{E}[\mathbf{X}_{l}] $ and $ \text{Var}[\mathbf{X}_{l}] $, which are denoted as  
\begin{equation}\label{eq16}
	\text{E}[\mathbf{X}_{l}] \leftarrow\mathbb{E}_\text{D}[\mu_{l}],  \text{Var}[\mathbf{X}_{l}]\leftarrow\frac{B}{B-1}\mathbb{E}_\text{D}[\sigma_l^2],
\end{equation}
where $ B $ denotes the mini-batch of the traing stage. $ \mathbb{E}_\text{D}[\cdot] $  denotes the expectation over the total training dataset. $ \mu_{l} $ and $ \sigma_l^2 $ denote  the mean and variance on the  mini-batch $ B $ for the input data of the BN layer at the  $l^{th}$ layer $\mathcal{C}\left(\cdot, \theta_l\right)$  of   HGNet, respectively.

Subsequently, $ \mathbf{X}_{l}^{\text{nor}} $ is transformed by the affine parameters to obtain the output of the BN layer at the  $l^{th}$ layer $\mathcal{C}\left(\cdot, \theta_l\right)$  of  HGNet, which is denoted as 
\begin{equation}\label{eq17}
 \mathbf{X}_{l}^{\text{bn}}[:,:,c]=\gamma_l^c\otimes \mathbf{X}_{l}^{\text{nor}}[:,:,c]+\beta_l^c,
\end{equation}
where  $ \mathbf{\gamma}_l= \left[\gamma_l^1,\cdots, \gamma_l^c,\cdots, \gamma_l^C  \right]\in \mathbb{R}^{C} $ and  $ \mathbf{\beta}_l= \left[\beta_l^1,\cdots, \beta_l^c,\cdots, \beta_l^C  \right]\in \mathbb{R}^{C} $ denote the affine parameters of the scale and the shift of the BN layer at  the  $l^{th}$ layer $\mathcal{C}\left(\cdot, \theta_l\right)$  of  HGNet, which  are learnable parameters. From Eqs.(\ref{eq15}) to (\ref{eq17}), $ \mathbf{X}_{l}^{\text{bn}} $ is determined by $ \mathbf{X}_{l} $,  $ \text{E}[\mathbf{X}_{l}] $,  $ \text{Var}[\mathbf{X}_{l}] $, $ \epsilon $, $ \mathbf{\gamma}_l $, $ \mathbf{\beta}_l $, where $ \text{E}[\mathbf{X}_{l}] $,  $ \text{Var}[\mathbf{X}_{l}] $ and $ \epsilon $ are constants after training. Therefore, $ \mathbf{X}_{l}^{\text{bn}} $ can be changed to improve the sum rate performance of the beamforming design by online adaptively updating $ \mathbf{\gamma}_l $ and $ \mathbf{\beta}_l $ in the inference phase.

On the other hand, updating $ \mathbf{\gamma}_l $ and $ \mathbf{\beta}_l $ also requires an objective function. This can be selected to the negative value of the sum rate as a loss function to update  $ \mathbf{\gamma}_l $ and $ \mathbf{\beta}_l $  via the gradient descent algorithm. Unfortunately, the computation of sum rate typically involves high-dimensional  matrix inverse operations, which hinders the realization of the online adaptive real-time beamforming design. As a remedy, this work uses information entropy as an objective function to optimize  $ \mathbf{\gamma}_l $ and $ \mathbf{\beta}_l $. This is because  the information entropy is highly computationally efficient, relying only on  simple dot product and summation operations without matrix inversion.  In addition, the information entropy can measure error and bias, which ensures to learn a better model \cite{bib37}.  Formally, this is denoted as
\begin{equation}
	\mathcal{L}_{\text{ie}}=-\sum_{q=1}^{Q_t}\sum_{i=1}^{I_t}\sum_{c=1}^{M}\left|\mathbf{V}_{t,\text{HGNet}}^{\text{3D}}[q,i,c]\otimes\log\mathbf{V}_{t,\text{HGNet}}^{\text{3D}}[q,i,c]\right|.
	\label{eq18}
\end{equation}

In summary, by minimizing $\mathcal{L}_{\text{ie}}$ to  adaptively update $ \mathbf{\gamma}_l $ and $ \mathbf{\beta}_l $ online, the proposed OAU algorithm effectively solves the above two problems. This also enables the online adaptive real-time beamforming design for dynamic wireless environments with the varying channels and the different numbers of APs and users. The pseudocode of the proposed OAU algorithm is summarized in Algorithm 1.

\begin{algorithm}[t]
	\caption{Proposed OAU  Algorithm}
	\label{alg3}
	\KwIn{Trained HGNet, number of layers of HGNet $L$, $ \mathbf{H}_t $, number of updates $ H $, learning rate $ R $\;}
	\KwOut{$\mathbf{V}_{t,\text{HGNet}}^{\text{3D}}$\;}
	Initialize  $\left( \text{E}[\mathbf{X}_{l}], \text{Var}[\mathbf{X}_{l}], \epsilon , \mathbf{\gamma}_l, \mathbf{\beta}_l  \right)$ $ \leftarrow $Trained HGNet\;
	\For{$h \leftarrow 1:H $}
	{ \For{$l \leftarrow 1:L $}{
		$\mathbf{X}_{l}^{\text{nor}}\leftarrow$ Calculate the BN normalization of $ \mathbf{H}_t $ at   the  $l^{th}$ layer $\mathcal{C}\left(\cdot, \theta_l\right)$ of HGNet by Eq.(\ref{eq15})\;
		$\mathbf{X}_{l}^{\text{bn}}\leftarrow$ Calculate the BN output of  $ \mathbf{H}_t $ at   the  $l^{th}$ layer $\mathcal{C}\left(\cdot, \theta_l\right)$  of HGNet by Eq.(\ref{eq17})\;		
	}
$\mathcal{L}_{\text{ie}}\leftarrow$ Calculate the loss function by Eq.(\ref{eq18})\;
 \While{$l>0 $}{
	$\mathbf{\gamma}_l \leftarrow \mathbf{\gamma}_l-R \times \partial \mathcal{L}_{\text{ie}} / \partial \mathbf{\gamma}_l $\;
	$\mathbf{\beta}_l \leftarrow \mathbf{\beta}_l-R \times \partial \mathcal{L}_{\text{ie}} / \partial \mathbf{\beta}_l $\;
	$l\leftarrow l-1$\;		
	}
}

$ \mathbf{V}_{t,\text{HGNet}}^{\text{3D}}\leftarrow $ Output beamforming by feeding $  \mathbf{H}_t $  into HGNet that have been online adaptively updated\;
	\textbf{Return:} $\mathbf{V}_{t,\text{HGNet}}^{\text{3D}}$\;
\end{algorithm} 

\section{Experimental Results}

In this section, we validate the effectiveness of the proposed HGNet with OAU algorithm. Specifically, we first introduce the experimental environments and system parameters. Then, we evaluate the performance of HGNet, followed by an assessment of the OAU algorithm’s performance. As benchmarks, the following schemes are compared:

$\bullet$ WMMSE: WMMSE \cite{ref9} is a classical traditional optimization algorithm for the beamforming design, which is frequently applied as a benchmark for comparing the beamforming design based on deep learning. 

$ \bullet $ Edge-GNN:  Based on GNNs, \cite{ref17} proposes Edge-GNN for cooperative beamforming design with better generalization performance. 

$\bullet$ SUNet: Based on CNNs, \cite{ref12} proposes SUNet with high computational efficiency to realize the beamforming design of cell-free systems. 

$ \bullet $ HGNet w/o  $\mathcal{G}\left(\cdot, \theta_l\right)$: This variant excludes the high-generalization beamforming module $\mathcal{G}\left(\cdot, \theta_l\right)$  to specifically assess the contribution of this module to overall performance.

\subsection{Experimental Setup} 
 
To characterize the varying channels of dynamic wireless environments, three commonly used channel models  have been selected as follows.

$\bullet$ Channel 1:  We utilize the geometric multi-path channel model \cite{bib22}, and the channel ${\bf{H}}_{i,t}^q\in\mathbb{C}^{M \times N}$ of the $q^{th}$ AP to  the $ i^{th} $ user at the $ t^{th} $ period can be expressed as
\begin{equation}
	{\bf{H}}_{i,t}^q = \beta_{i,t}^q \sum_{p=1}^{P}\frac{g^q_{i,t,p}}{\sqrt{P}}
	{\bf{a}}_r\left(\phi^q_{i,t,p}\right){\bf{a}}_t^{{H}}\left(\varphi^q_{i,t,p}\right),
	\label{eq22}
\end{equation}
where $\beta_{i,t}^q$ denotes the large-scale fading coefficient of the $q^{th}$ AP to  the $ i^{th} $ user at the $ t^{th} $ period. $P$ denotes the number of propagation paths. $g^q_{i,t,p}\sim\mathcal{CN}\left(0,1\right)$, $\varphi^q_{i,t,p}$ and $\phi^q_{i,t,p}$ denote the complex path gain, angles of departure and arrival, respectively. Moreover, ${\bf{a}}_t$ and ${\bf{a}}_r$ denote the array responses of  transceiver.

For other types of  channels, we  adopt the channel model in \cite{bib23} as follows
\begin{equation}
	{\bf{H}}_{i,t}^q = \beta_{i,t}^q \left(
	\sqrt{\frac{\epsilon}{\epsilon+1}}{\bf{a}}_t\left(\varphi^q_{i,t}\right)
	{\bf{a}}_r^{{H}}\left(\phi^q_{i,t}\right)+\sqrt{\frac{1}{\epsilon+1}}  {\bar{\bf{H}}}_{i,t}^q
	\right),
	\label{eq20}
\end{equation}
where $\varphi^q_{i,t}$ and $\varphi^q_{i,t}$
denote the LoS directions of transmit and receive, respectively. ${\bar{\bf{H}}}_{i,t}^q$ 
denotes the NLoS component following the complex Gaussian distribution $ \mathcal{CN}\left({\bf{0}},{\bf{I}}\right)$.

$\bullet$ Channel 2: Rayleigh channel with
$\epsilon = 0$.

$\bullet$ Channel 3: Rice channel with $\epsilon = 3\, {\rm{dB}}$.

\begin{figure}[t]
	\centering
	\includegraphics[scale=0.60]{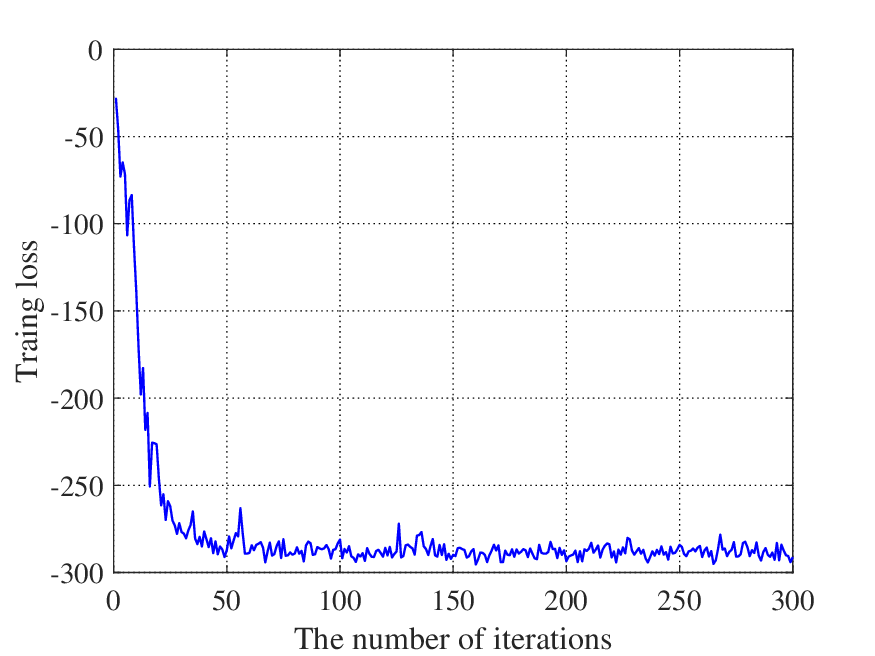}
	\caption{Plot of training loss with the number of iterations.}
	\label{fig2}
\end{figure}

The proposed HGNet with OAU algorithm is implemented by PyTorch. For HGNet, the  Adam optimizer is selected during the training stage. The learning rate and batch size are set to 64 and 0.1, respectively. The number of layers $ L $ for  HGNet is set to 5, where  the architectural parameters of each convolution unit $\mathcal{C}\left(\cdot, \theta_l\right)$ are set to  $k_l^w=3$, $k_l^h=3$,  $p_l^w=1$, $p_l^h=1$, $s_l^w=1$, $s_l^h=1$, $c_l=2M$ to satisfy \emph{Propositions 1}  or \emph{2}. By randomly scattering $ Q=16 $ APs with $ M=4 $ antennas and $ I=16 $ users with $ N=2 $ antennas in $ 500 \times 500 (m^2) $ size areas, each channel model generates 6400 training data, where each channel also generates 640 test data to verify the generalization performance of the trained  HGNet. For better training, a total of 19200 data for the three channels are randomly scrambled and fed into HGNet by utilizing the DataLoader database in Python. In addition, the plot of training loss with the number of iterations is shown in Fig.\ref{fig2}. For OAU algorithm, the optimizer is also selected the  Adam, where the learning rate  $ R $ is set to 0.001. 

\subsection{Experimental Results for HGNet}

\begin{figure}[t]
	\centering
	\includegraphics[scale=0.60]{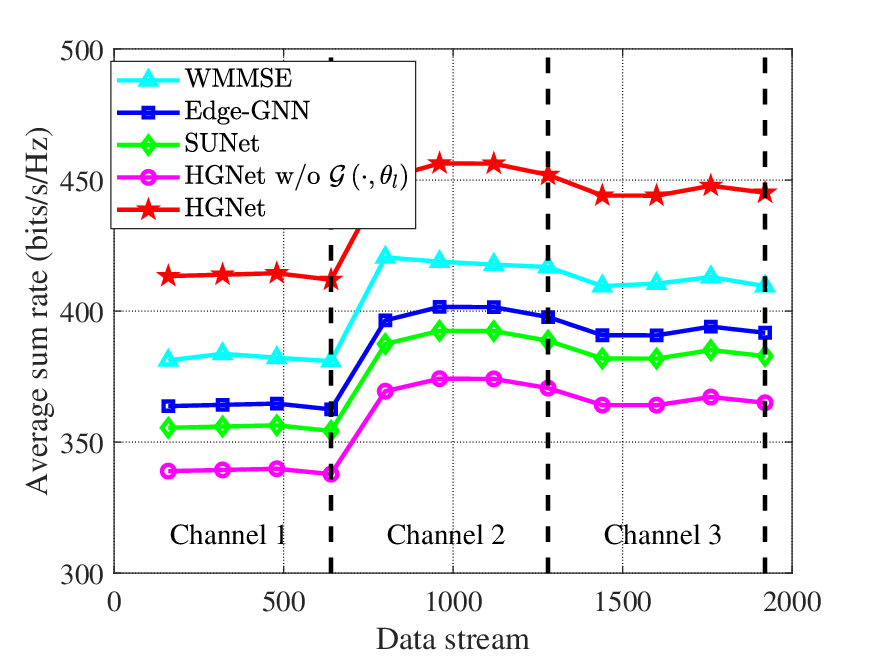}
	\caption{Comparative results of average generalization sum rate across three channels under $Q_t=24$, $I_t=24$.}
	\label{fig3}
\end{figure}

\begin{figure}[t]
	\centering
	\includegraphics[scale=0.60]{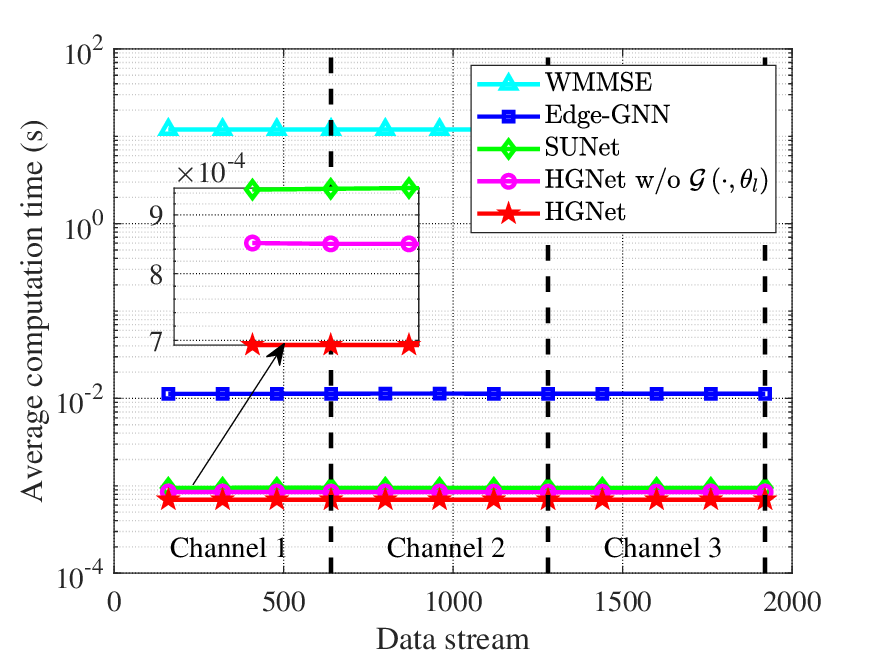}
	\caption{Comparative results of average computation time across three channels under  $Q_t=24$, $I_t=24$.}
	\label{fig4}
\end{figure}

\begin{figure}[t]
	\centering
	\includegraphics[scale=0.60]{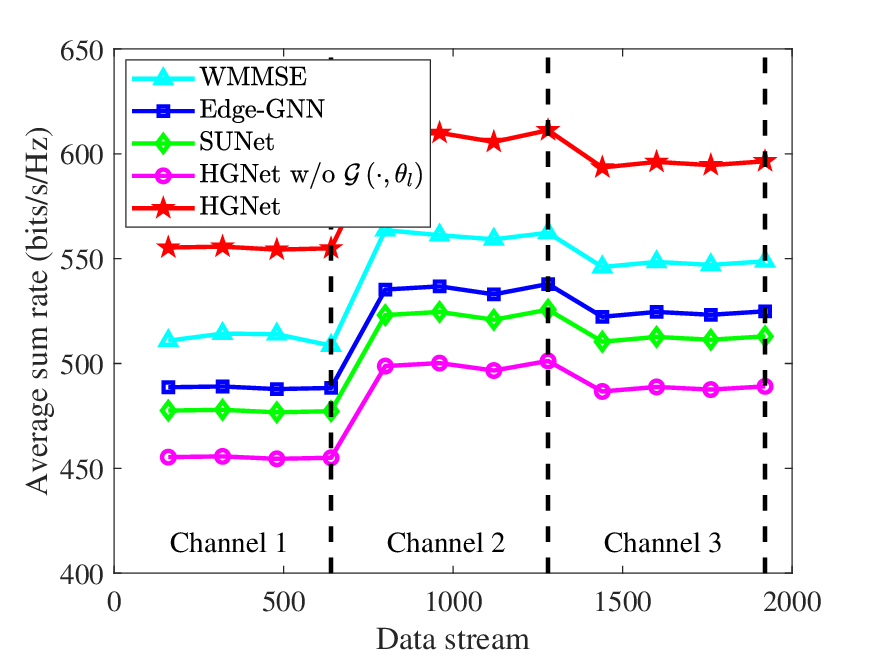}
	\caption{Comparative results of  average generalization sum rate across three channels under $Q_t=32$, $I_t=32$.}
	\label{fig5}
\end{figure}

\begin{figure}[t]
	\centering
	\includegraphics[scale=0.60]{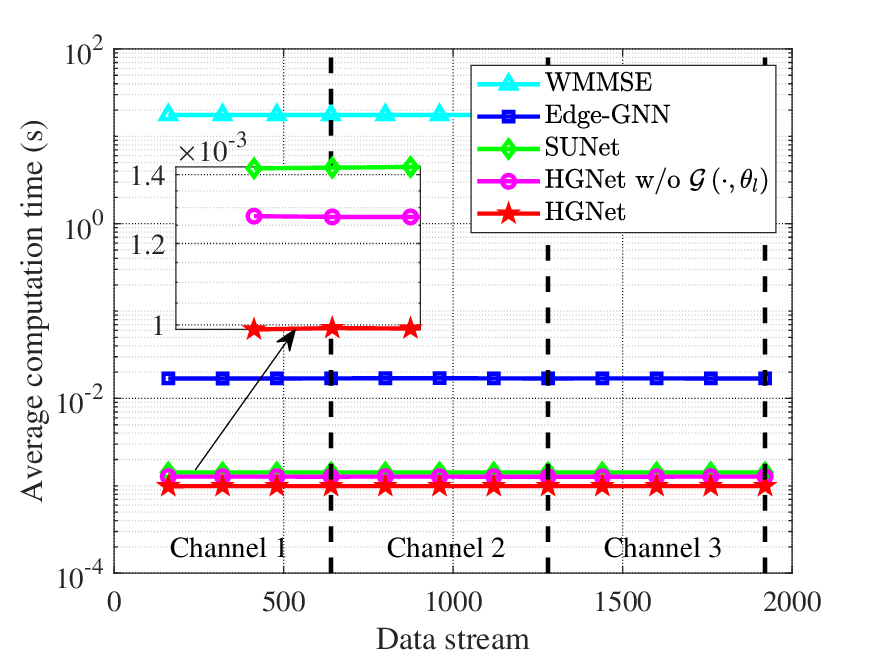}
	\caption{Comparative results of average computation time across three channels under $Q_t=32$, $I_t=32$.}
	\label{fig6}
\end{figure}

To verify the effectiveness of the proposed HGNet on the varying channels and different numbers of APs and users, the average generalization sum rate and the average  computation time of these compared algorithms are shown in Figs.\ref{fig3}-\ref{fig6}, respectively. For  Edge-GNN, SUNet, HGNet and HGNet w/o $\mathcal{G}\left(\cdot, \theta_l\right)$, the number of APs and users  is fixed at $Q_t=16, I_t=16$ in the training stage.  These trained models are then generalized to scenarios with  $Q_t=24, I_t=24$ and $Q_t=32, I_t=32$ during the inference stage, respectively. Since WMMSE is a traditional optimization algorithm without the training and inference stages, it is applied directly in  $Q_t=24, I_t=24$ and $Q_t=32, I_t=32$ scenarios.  By comparing the average generalization sum rate  of  HGNet and HGNet w/o $\mathcal{G}\left(\cdot, \theta_l\right)$ in Figs.\ref{fig3} and \ref{fig5}, we observe that the high-generalization beamforming module $\mathcal{G}\left(\cdot, \theta_l\right)$ effectively improves the generalization sum rate performance of the varying channels. This is because the high-generalization beamforming module $\mathcal{G}\left(\cdot, \theta_l\right)$ effectively extracts the valuable features for the varying channels. In addition, as can be seen from  Figs.\ref{fig4} and \ref{fig6}, the average
computation time of HGNet  is also lower than that of HGNet w/o $\mathcal{G}\left(\cdot, \theta_l\right)$, since the module $\mathcal{G}\left(\cdot, \theta_l\right)$ discards some sensitive features for the varying channels, thereby reducing  computational complexity.

By comparing the experimental results of HGNet with WMMSE, SUNet and Edge-GNN in Figs.\ref{fig3}-\ref{fig6}, HGNet achieves a better sum rate with lower computation time for the varying channels with different numbers of APs and users. The reasons are as follows. Although WMMSE is a traditional optimization algorithm for the beamforming design, it has the highest computational complexity due to multiple iterations and matrix inversions. Compared to WMMSE, Edge-GNN reduces the average computation time by utilizing GNNs to improve the computational efficiency, but it is designed for $D\left(\mathbf{H}_t \right) = D\left(\mathbf{H}_{t^\prime} \right), \forall (t\neq t^\prime)\in\mathcal{T}$. When encountering $D\left(\mathbf{H}_t \right) \neq D\left(\mathbf{H}_{t^\prime} \right), \forall (t\neq t^\prime)\in\mathcal{T}$, the generalization sum rate performance of Edge-GNN is degraded, since the deep learning assumption that the training and the inference stages have the same distribution for better generalization performance is violated. Similarly, SUNet is also designed for $D\left(\mathbf{H}_t \right) = D\left(\mathbf{H}_{t^\prime} \right), \forall (t\neq t^\prime)\in\mathcal{T}$, where the generalization  sum rate performance also degrades in $D\left(\mathbf{H}_t \right) \neq D\left(\mathbf{H}_{t^\prime} \right), \forall (t\neq t^\prime)\in\mathcal{T}$. However, SUNet further improves the computational efficiency  by  using the weight sharing mechanism of CNNs. On the contrary, in addition to utilizing convolution units to improve the computational efficiency,  HGNet designs the high-generalization beamforming module $\mathcal{G}\left(\cdot, \theta_l\right)$ to extract the valuable features for $D\left(\mathbf{H}_t \right) \neq D\left(\mathbf{H}_{t^\prime} \right), \forall (t\neq t^\prime)\in\mathcal{T}$, which effectively improves the generalization sum rate performance of the  varying channels. Consequently, HGNet achieves a higher sum rate with  a lower computation time in the order of $10^{-3}$ seconds, which implements the real-time beamforming design for dynamic wireless environments with the varying channels and the different numbers of APs and users in cell-free systems.

\begin{figure}[t]
	\centering
	\includegraphics[scale=0.60]{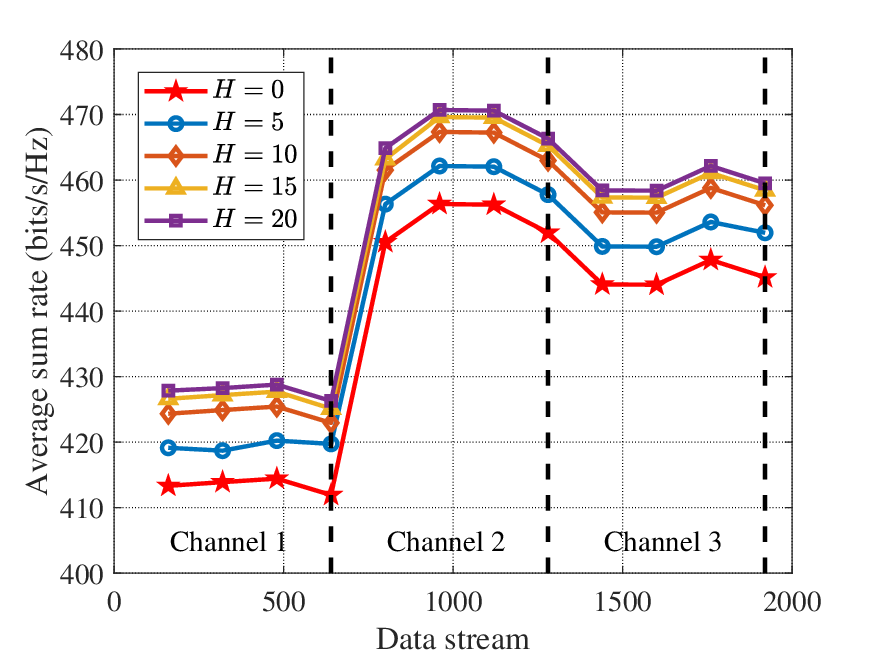}
	\caption{Average generalization  sum rate of OAU algorithm across three channels under $Q_t=24$, $I_t=24$.}
	\label{fig7}
\end{figure} 

\begin{figure}[t]
	\centering
	\includegraphics[scale=0.60]{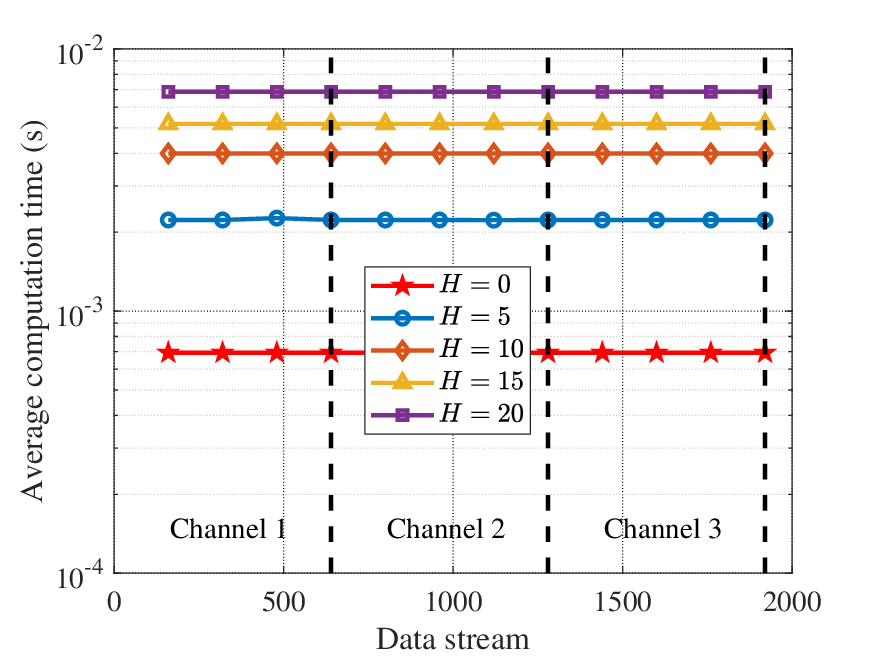}
	\caption{Average computation  time of OAU algorithm across three channels under $Q_t=24$, $I_t=24$.}
	\label{fig8}
\end{figure}

\begin{figure}[t]
	\centering
	\includegraphics[scale=0.60]{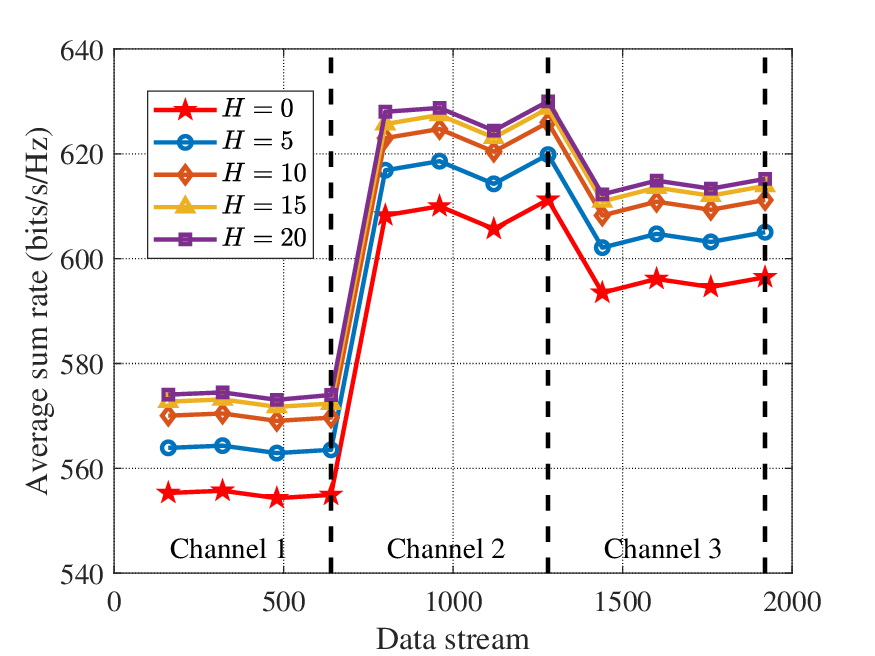}
	\caption{Average generalization  sum rate of OAU algorithm across three channels under $Q_t=32$, $I_t=32$.}
	\label{fig9}
\end{figure}

\begin{figure}[t]
	\centering
	\includegraphics[scale=0.60]{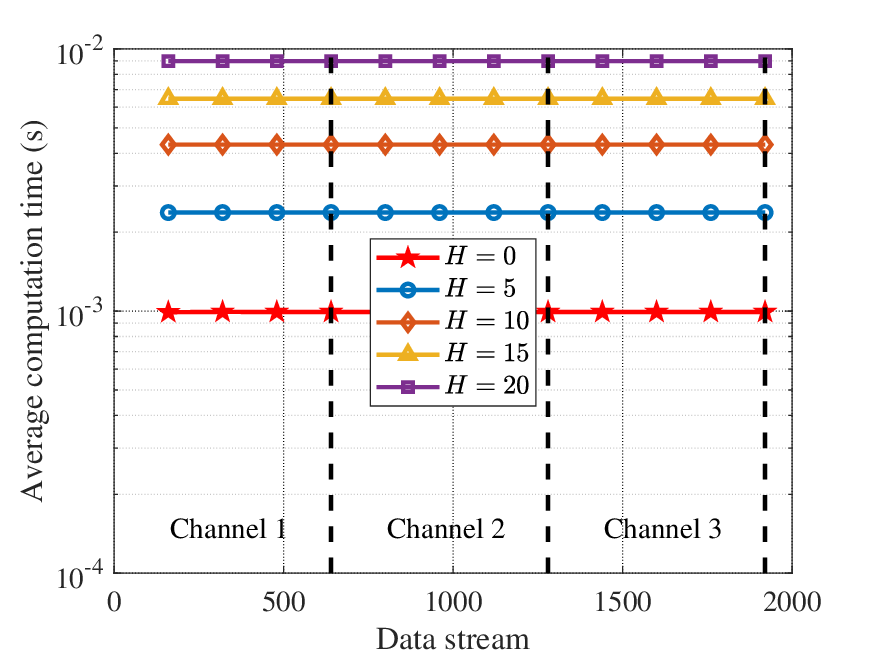}
	\caption{Average computation  time of OAU algorithm across three channels under $Q_t=32$, $I_t=32$.}
	\label{fig10}
\end{figure}

\subsection{Experimental Results for OAU  Algorithm}

To verify the effectiveness of the proposed OAU  algorithm on the varying channels and the different numbers of APs and users, the average generalization sum rate and the average computation time at different number of iterations $H$ are shown in Figs.\ref{fig7}-\ref{fig10}, respectively.  When the number of iterations  $H$ is set to 0, the proposed OAU algorithm is not applied to update the parameters of HGNet. This is used as a comparative baseline for  other different numbers of iterations $H$. As can be seen from  Figs.\ref{fig7} and \ref{fig9}, the  average generalization sum rate gradually increases as $H$ rises from 0 to 15, and does not increase significantly from 15 to 20. This improvement occurs because the proposed OAU algorithm  iteratively updates the parameters of HGNet to improve the sum rate performance. As can be seen from  Figs.\ref{fig8} and \ref{fig10}, with increasing the number of iterations $H$ from 0 to 20,  the average computation time becomes progressively higher compared to the number of iterations $H=0$. This is attributed to the fact that the proposed OAU algorithm consumes time to update the parameters of HGNet. Although the proposed OAU algorithm increases the sum rate performance come at the expense of computation time, the average computation time remains below  $10^{-2}$ seconds, i.e., on the order of milliseconds. This  meets the requirements of the real-time beamforming design.

\section{Conclusion}

In this paper,  HGNet with  OAU algorithm is proposed to enable the online adaptive real-time beamforming design for dynamic wireless environments with the varying channels and the different numbers of APs and users in cell-free systems. HGNet utilizes the residual structure of CNNs to adapt to the varying numbers of  APs and users. Meanwhile,  HGNet designs the high-generalization beamforming module to extract the valuable features of the varying channels for improving the generalization sum rate performance. Moreover, the OAU algorithm provides an online adaptive update mechanism for HGNet’s parameters, enabling  the online adaptive real-time beamforming design. Numerical results show that the proposed HGNet with OAU algorithm achieves a higher sum rate with a computation burden in the order of milliseconds, effectively meeting the demands of the real-time beamforming of cell-free systems in dynamic wireless environments.

\section*{Appendix A}
\section*{Proof of Proposition 3}
To facilitate the analysis of the generalization error of $\mathcal{G}\left(\cdot, \theta_l\right)$ in HGNet, let $ R_{\mathcal{T}_{\text{in}}}\left[\mathcal{G},\mathcal{G}^\prime\right]=\underset{x\sim  \mathcal{T}_{\text{in}}}{\mathbb{E}}\left[ \ell\left(\mathcal{G}\left(x, \theta_l\right),\mathcal{G}^\prime\left(x, \theta_l^\prime\right)\right)\right]  $ denote the generalization error risk of a pair $ \mathcal{G}\left(\cdot, \theta_l\right) $ and $ \mathcal{G}^\prime\left(\cdot, \theta_l^\prime\right) $ on the target domain $ \mathcal{T}_{\text{in}} $, where $ \ell\left( \cdot\right)  $ denotes a error function. Note that the generalization error risk of $\mathcal{G}\left(\cdot, \theta_l\right)$ on $ \mathcal{T}_{\text{in}} $ is $ R_{\mathcal{T}_{\text{in}}}\left[\mathcal{G},0\right]=\underset{x\sim  \mathcal{T}_{\text{in}}}{\mathbb{E}}\left[ \ell\left(\mathcal{G}\left(x, \theta_l\right),0\right)\right]  $, and $ R_{\mathcal{T}_{\text{in}}}\left[\mathcal{G}\right]=R_{\mathcal{T}_{\text{in}}}\left[\mathcal{G},0\right] $ for simplification. Moreover, let $\mathcal{G}^*\left(\cdot, \theta_l^*\right)$ denote the ideal $\mathcal{G}\left(\cdot, \theta_l\right)$.  By utilizing the triangular inequality, the following inequality can be obtained as
\begin{equation}\label{key1}
R_{\mathcal{T}_{\text{in}}}\left[\mathcal{G}\right]\leq R_{\mathcal{T}_{\text{in}}}\left[\mathcal{G^{*}}\right]+R_{\mathcal{T}_{\text{in}}}\left[\mathcal{G^{*}},{\mathcal{G}}\right]+R_{\mathcal{\overline{\mathcal{T}_{\text{in}}}}}\left[\mathcal{G^{*}},{\mathcal{G}}\right]-R_{\mathcal{\overline{T}_{\text{in}}}}\left[\mathcal{G^{*}},{
	\mathcal{G}}\right],	
\end{equation} 
where $ \overline{\mathcal{T}_{\text{in}}}=\sum_{t=1}^{T}\kappa_{t}\mathcal{S}_{\text{tr}}^t$ denotes a mixture of $ T $ varying channels closest to the target domain $ \mathcal{T}_{\text{in}} $ with $ \sum_{t=1}^{T} \kappa_{t}=1 $. 
By bringing $ \overline{\mathcal{T}_{\text{in}}}=\sum_{t=1}^{T}\kappa_{t}\mathcal{S}_{\text{tr}}^t$ to Eq.(\ref{key1}),
\begin{equation}
\begin{aligned}\label{key2}
R_{\mathcal{T}_{\text{in}}}\left[\mathcal{G}\right]\leq& R_{\mathcal{T}_{\text{in}}}\left[\mathcal{G^{*}}\right]+\sum_{t=1}^{T}\kappa_{t}R_{\mathcal{S}_{\text{tr}}^t}\left[\mathcal{G^{*}}, {\mathcal{G}}\right]\\&+\sum_{t=1}^{T}\kappa_{t}R_{\mathcal{T}_{\text{in}}}\left[\mathcal{G^{*}}{,\mathcal{G}}\right]-\sum_{t=1}^{T}\kappa_{t}R_{\mathcal{S}_{\text{tr}}^t}\left[\mathcal{G^{*}}{,\mathcal{G}}\right],
\end{aligned}
\end{equation}

\emph{Lemma 1: (Lemma 5.3  in  \cite{bib24}) With the previous	definitions 1 and 2, the following inequality is held as $$ R_{\mathcal{T}_{\text{in}}}\left[\mathcal{G^{*}}{,\mathcal{G}}\right] \leq R_{\mathcal{S}_{\text{tr}}^t}\left[\mathcal{G^{*}}{,\mathcal{G}}\right] +d_{\text{G-MMD}}\left({\mathcal{S}_{\text{tr}}^t},{\mathcal{T}_{\text{in}}}\right).$$}

Based on \emph{Lemma 1}, Eq.(\ref{key2}) is rewritten as
\begin{equation}
	\begin{aligned}\label{key3}
		R_{\mathcal{T}_{\text{in}}}\left[\mathcal{G}\right]\leq& R_{\mathcal{T}_{\text{in}}}\left[\mathcal{G^{*}}\right]+\sum_{t=1}^{T}\kappa_{t}R_{\mathcal{S}_{\text{tr}}^t}\left[\mathcal{G^{*}}, {\mathcal{G}}\right]\\&+\sum_{t=1}^{T}\kappa_{t}d_{\text{G-MMD}}\left({\mathcal{S}_{\text{tr}}^t},{\mathcal{T}_{\text{in}}}\right),
	\end{aligned}
\end{equation}
where $ \sum_{t=1}^{T}\kappa_{t}R_{\mathcal{S^{\mathit{t}}}}\left[\mathcal{G^{*}}, {\mathcal{G}}\right] $  and $ \sum_{t=1}^{T}\kappa_{t}d_{\text{G-MMD}}\left({\mathcal{S}_{\text{tr}}^t},{\mathcal{T}}_{\text{in}}\right) $  are applied to the triangular inequality, i.e., 
\begin{equation}
	\begin{aligned}\label{key4}
		\sum_{t=1}^{T}\kappa_{t}R_{\mathcal{S}_{\text{tr}}^t}\left[\mathcal{G^{*}}, {\mathcal{G}}\right]&\leq\sum_{t=1}^{T}\kappa_{t}R_{\mathcal{S}_{\text{tr}}^t}\left[\mathcal{G^{*}}\right]+\sum_{t=1}^{T}\kappa_{t}R_{\mathcal{S}_{\text{tr}}^t}\left[ {\mathcal{G}}\right]\\&\leq R_{\overline{\mathcal{T}}}\left[\mathcal{G^{*}}\right]+\sum_{t=1}^{T}\kappa_{t}R_{\mathcal{S}_{\text{tr}}^t}\left[ {\mathcal{G}}\right],
	\end{aligned}
\end{equation}
\begin{equation}
	\begin{aligned}\label{key5}
		 &\sum_{t=1}^{T}\kappa_{t}d_{\text{G-MMD}}\left({\mathcal{S}_{\text{tr}}^t},{\mathcal{T}}_{\text{in}}\right)\\&\leq\sum_{t=1}^{T}\kappa_{t}d_{\text{G-MMD}}\left(\mathcal{S}_{\text{tr}}^{\mathit{t}},\overline{\mathcal{T}_\text{in}},\right)+\sum_{t=1}^{T}\kappa_{t}d_{\text{G-MMD}}\left(\overline{\mathcal{T}_\text{in}},\mathcal{T}_\text{in}\right)\\&\leq\underset{t,t^{\prime}\in\mathcal{T}}{\textrm{sup}}d_{\text{G-MMD}}\left(\mathcal{S}_{\text{tr}}^{\mathit{t^{\prime}}},\mathcal{S}_{\text{tr}}^{\mathit{t}}\right)+d_{\text{G-MMD}}\left(\overline{\mathcal{T}_\text{in}},\mathcal{T}_\text{in}\right).
	\end{aligned}
\end{equation}
By bringing Eqs.(\ref{key4}) and (\ref{key5}) into Eq.(\ref{key3}),
\begin{equation}
	\begin{aligned}\label{key6}
		R_{\mathcal{T}}\left[\mathcal{G}\right]\leq&\underset{t,t^{\prime}\in\mathcal{T}}{\textrm{sup}}d_{\text{G-MMD}}\left(\mathcal{S}_{\text{tr}}^{\mathit{t^{\prime}}},\mathcal{S}_{\text{tr}}^{\mathit{t}}\right)+d_{\text{G-MMD}}\left(\overline{\mathcal{T}_\text{in}},\mathcal{T}_\text{in}\right)\\&+\varpi+\varrho,
	\end{aligned}
\end{equation}
where $ \varpi= \sum_{t=1}^{T}\kappa_{t}R_{\mathcal{S}_{\text{tr}}^t}\left[ {\mathcal{G}}\right] $  denotes the mixture weight of the generalization error for known source domains  $ \mathcal{S}_{\text{tr}}^t , t\in \mathcal{T}$. $ \varrho= R_{\overline{\mathcal{T}}}\left[\mathcal{G^{*}}\right]+R_{\mathcal{T}}\left[\mathcal{G^{*}}\right] $ denotes the combined error of ideal $\mathcal{G}^*\left(\cdot, \theta_l^*\right)$.

\emph{Lemma 2: (Theorem 29  in  \cite{bib25}) With the previous definitions 1 and 2, let $ \hat{\overline{\mathcal{T}_\text{in}}} $ and   $ \hat{{\mathcal{T}_\text{in}}} $ denote the $ D $ data sampled from $ \overline{\mathcal{T}_\text{in}} $ and $ \mathcal{T}_\text{in} $, respectively. Then for all $ \delta \in (0,1) $, with probability at least $ 1-\delta $,	the following inequality is held as
	\begin{align*}
		&d_{\text{G-MMD}}\left(\overline{\mathcal{T}_\text{in}},\mathcal{T}_\text{in}\right)\leq d_{\text{G-MMD}}\left(\hat{\overline{\mathcal{T}_\text{in}}},\hat{\mathcal{T}_\text{in}}\right)+2\sqrt{\frac{\log\left(\frac{2}{\delta}\right)}{2D}}\\&+\frac{2}{D}\left(\sum_{t=1}^{T}\kappa_{t}{\mathbb{E}}\left[\sqrt{\text{tr}\left(k_{{\hat{\mathcal{S}}}_{\text{tr}}^{t}}\right)}\right]+{\mathbb{E}}\left[\sqrt{\text{tr}\left(k_{\hat{\mathcal{T}_{\text{in}}}}\right)}\right]\right),
	\end{align*}
where  $ k_{{\hat{\mathcal{S}}}_{\text{tr}}^{t}} $ and $ k_{\hat{\mathcal{T}_{\text{in}}}} $  denote kernel functions computed on samples from  $ \hat{\mathcal{S}}_{\text{tr}}^t$ and  $\hat{\mathcal{T}_\text{in}} $, respectively.}

By bringing \emph{Lemma 2} into Eq.(\ref{key6}), we complete the proof of \emph{Proposition 3}.

\bibliographystyle{IEEEtran}
\bibliography{ref.bib}

\end{document}